\documentclass[fleqn,11pt]{wlscirep}

\usepackage{float}
\usepackage{framed}
\usepackage{ulem}
\usepackage{bm}
\usepackage{physics}
\usepackage{amsfonts}
\usepackage[utf8]{inputenc}
\usepackage{multibib}

\usepackage{textcomp}
\usepackage[T1]{fontenc}
\usepackage{mathpazo} % get Palatino math  
\usepackage{amsmath}
\usepackage{gensymb}
\usepackage{xcolor}
\newcommand{\rev}[1]{\textcolor{black}{#1}}
\newcommand{\rrev}[1]{\textcolor{black}{#1}}

\title{Spin-glass states generated in \rrev{a} van der Waals magnet by alkali-ion intercalation}

\author[1,*]{S. Khan}
\author[2]{E. S. Y. Aw}
\author[3]{L. A. V. Nagle-Cocco}
\author[4]{A. Sud}
\author[5]{S. Ghosh}
\author[2]{M. K. B. Subhan}
\author[1]{Z. Xue}
\author[1]{C. Freeman}
\author[1]{D. Sagkovits}
\author[6]{A. Guti\'errez-Llorente}
\author[7]{I. Verzhbitskiy}
\author[8]{D. M. Arroo}
\author[1]{C. W. Zollitsch}
\author[7]{G. Eda}
\author[5,9,10]{E. J. G. Santos}
\author[3]{S. E. Dutton}
\author[1]{S. T. Bramwell}
\author[2]{C. A. Howard}
\author[1,4,11,+]{H. Kurebayashi}
\affil[1]{London Centre for Nanotechnology, University College London, London, WC1H 0AH, United Kingdom}
\affil[2]{Department of Physics $\&$ Astronomy, University College London, London, WC1H 0AH, United Kingdom}
\affil[3]{Cavendish Laboratory, University of Cambridge, Cambridge, CB3 0HE, United Kingdom}
\affil[4]{WPI-AIMR, Tohoku University, 2-1-1, Katahira, Sendai 980-8577, Japan}
\affil[5]{Institute for Condensed Matter Physics and Complex Systems, School of Physics and Astronomy, University of Edinburgh, Edinburgh EH9 3FD, United Kingdom}
\affil[6]{Escuela Superior de Ciencias Experimentales y Tecnología, Universidad Rey Juan Carlos, Madrid 28933, Spain}
\affil[7]{Physics Department, National University of Singapore, Singapore, Singapore.}
\affil[8]{Department of Materials, Imperial College London, Exhibition Road, London, SW7 2AZ, United Kingdom}
\affil[9]{Higgs Centre for Theoretical Physics, The University of Edinburgh,  EH9 3FD, United Kingdom}
\affil[10]{Donostia International Physics Center, 20018 Donostia-San Sebastián, Basque Country, Spain}
\affil[11]{Department of Electronic and Electrical Engineering, University College London, London, WC1E 7JE, United Kingdom}

\affil[*]{safe.khan.11@ucl.ac.uk}
\affil[+]{h.kurebayashi@ucl.ac.uk}

\begin{abstract}
Tuning magnetic properties in layered van der Waals (vdW) materials has captured a significant attention due to the efficient control of ground-states by heterostructuring and external stimuli. Electron doping by electrostatic gating, interfacial charge transfer and intercalation is particularly effective in manipulating the exchange and spin-orbit properties, resulting in a control of Curie temperature ($T_{\text{C}}$) and magnetic anisotropy. Here, we discover an uncharted role of intercalation to generate magnetic frustration. As a model study, we intercalate Na atoms into the vdW gaps of pristine Cr$_2$Ge$_2$Te$_6$ (CGT) where generated magnetic frustration leads to emerging spin-glass states coexisting with a ferromagnetic order. A series of dynamic magnetic susceptibility measurements/analysis confirms the formation of magnetic clusters representing slow dynamics with a distribution of relaxation times. The intercalation also modifies other macroscopic physical parameters including the significant enhancement of $T_{\text{C}}$ from 66\,K to 240\,K and the switching of magnetic easy-hard axis direction. Our study identifies intercalation as a unique route to generate emerging frustrated spin states in simple vdW crystals.  
\end{abstract}

\begin{document}

\flushbottom
\maketitle

\thispagestyle{empty}
\newpage
\section{Introduction}

The discovery of long-range magnetic order in layered van der Waals (vdW) materials down to individual monolayers~\cite{HuangNature2017,Gong_Nature2017} has opened up new avenues for improving the understanding of fundamental concepts in condensed matter physics~\cite{bedoya2021intrinsic, gibertini2019magnetic,wang2022magnetic,kosterlitz1978phase,de1990magnetic,bramwell1993magnetization,van1997study,irkhin1999kosterlitz,cortie2020two,Tang_PhysRep2023}, together with potential for future applications~\cite{elahi2022review,kurebayashi2022MagnetismSOT,zhang2021two, lan2020magneto,khela2023laser}. For instance, a clean interface can be achieved in heterostructures made out of vdW materials and these devices may then be deployed for efficient switching of magnetisation state by spin-orbit torques~\cite{elahi2022review,kurebayashi2022MagnetismSOT}, as well as for magneto-optical applications~\cite{zhang2021two, lan2020magneto,khela2023laser}. 

Among the vdW magnetic materials, the families of chromium halides (i.e. CrI$_{3}$, CrBr$_{3}$ and CrCl$_{3}$), lamellar ternary chalcogenides (i.e. Cr$_2$Ge$_2$Te$_6$ and Cr$_2$Si$_2$Te$_6$), metal thiophosphates (MPS$_3$, M = Fe, Mn and Ni) and iron-based tellurides (i.e. Fe$_3$GeTe$_2$ and Fe$_5$GeTe$_2$), have been intensively and extensively studied~\cite{khan2020recentreview, hossain2022synthesis,ningrum2020recent,yan2022layer,neal2020symmetry}. One common aspect shared by most of these materials is the magnetic ordering transition temperature, Curie temperature ($T_{\text{C}}$) for ferromagnets, lying below room temperature~\cite{kurebayashi2022MagnetismSOT}, with a few exceptions including  Fe$_5$GeTe$_2$~\cite{kurebayashi2022MagnetismSOT} and Fe$_3$GaTe$_2$~\cite{Zhang_NComm2022}. In order to fully realise the potential of 2D magnetic materials for practical applications, a consequential drive is taking place in either discovering new materials with above-room-temperature $T_{\text{C}}$, or to elevate $T_{\text{C}}$ of existing systems by external means. For the latter, taking Cr$_2$Ge$_2$Te$_6$ (CGT) as an example, its $T_{\text{C}}$ of 66\,K~\cite{Khan2019SpinDynamicsPRB, zhang2016magnetic} is reported to be significantly enhanced by electrostatic gating~\cite{IvanVerzhbitskiy2020Controlling,Zhuo_AdvMater2021}, strain~\cite{Bhoi_PRL2021,O’Neill_ACSNano2023} and intercalation~\cite{Wang_JACS2019}. 

The modification of magnetic properties in vdW systems has been primarily focused on parameters such as $T_{\text{C}}$ and magnetic anisotropy constants that arise from the exchange and spin-orbit interaction respectively. Electron doping alters the type and strength of the exchange interaction~\cite{Zener_RMP1953,Matsukura_NNano2015}, e.g. via the change of the Fermi level position for itinerant ferromagnets~\cite{Deng_Nature2018}, and the number of carriers that play a leading role in carrier-mediated indirect Ruderman–Kittel–Kasuya–Yosida (RKKY) and double-exchange mechanisms for less-conductive magnets~\cite{IvanVerzhbitskiy2020Controlling}. Doped electrons populate in electronic bands with a specific orbital character, modifying the spin-orbit property impacting on magnetic anisotropies~\cite{Bruno_PRB1989}. These material properties represent the homogeneous nature of systems. By contrast, some important yet unanswered questions in researching vdW magnetic systems are: what arise when a vdW magnetic system is introduced with inhomogeneous carrier-doping, and what are resultant parameter changes that we can experimentally access to? To address this with a model case, we employ Na-ion intercalation into the vdW gaps of pristine Cr$_2$Ge$_2$Te$_6$ where Na ions are inhomogeneously intercalated within the gaps. We actively use this nature to intentionally create disorder and study its consequence to magnetism.

While inhomogeneity might be considered negative since it breaks the periodicity/coherency of crystallinity in real space, it can however generate new physics that cannot be achieved otherwise. In magnetism, spin-glass states~\cite{Cannella_PRB1972} emerge because of the presence of multiple local energy minima that the system freezes into, generated by inhomogeneity and frustration/competition between the ferromagnetic and anti-ferromagnetic exchange interactions. Hopping between the energy minima originates experimentally measurable magnetic responses that are time-, temperature- and magnetic-field-dependent~\cite{Mydosh1993SpinGlassBook}. Characteristic of the phase transition, the spin-glass has an order parameter defined by Edwards and Anderson~\cite{Edwards_1975}, together with the critical temperature of its freezing, $T_\text{g}$. The research field of spin-glasses is vast and diverse, ranging from canonical spin-glass states often observed in non-magnetic materials with randomly distributed magnetic impurities~\cite{Mydosh2015SpinGlassReviewPaper} to more complex spin-glassy behaviours in frustrated ferromagnets, called reentrant spin-glasses ~\cite{Gabay_PRL1981}. Spin glass formation in highly correlated systems is also known to herald yet more exotic behaviour such as high-temperature superconductivity in the cuprates~\cite{Chou_PRL1995}. The realisation of glassy magnetic states in vdW materials therefore suggests their significant potential for supporting new magnetic and electronic states, formed by tuning material properties via substitution, intercalation and exfoliation or heterostructuring. \rev{However, there are a limited number of reported studies on spin-glass states in vdW systems and all of them use magnetic disorders and frustrations arising in their host crystals~\cite{Takano_JAP2003,MASUBUCHI_JAC2008,GOOSSENS_JMMM2013,McGuire_PRMater2018}.}

In this study, we intercalated pristine CGT single crystals with alkali (sodium, Na)-ions to create magnetically-disordered Na-CGT. As a result, we discovered a frustrated magnetic state in the vicinity of 35\,K by dynamic magnetic susceptibility measurements suggesting the nature of this transition to be emergence of spin-glassiness behaviour. Further quantitative analysis points towards Na-CGT hosting a cluster spin-glass state at low temperatures, a feature not present in pristine CGT. This emergent frustrated state is accompanied by a clear increase of electric conductivity, enhanced Tc to $\sim 240$\,K and the sign reversal of two-fold magnetic anisotropy term that removes its out-of-plane magnetic easy axis and creates the magnetic easy plane in Na-CGT. Our study illustrates that vdW magnetic systems are highly tunable, not only for the exchange and spin-orbit coupling properties, but also now for magnetic frustration by intercalation techniques.\\

\section{Results $\&$ Discussion}

\begin{figure}[h!]
\centering
\includegraphics[width=18cm]{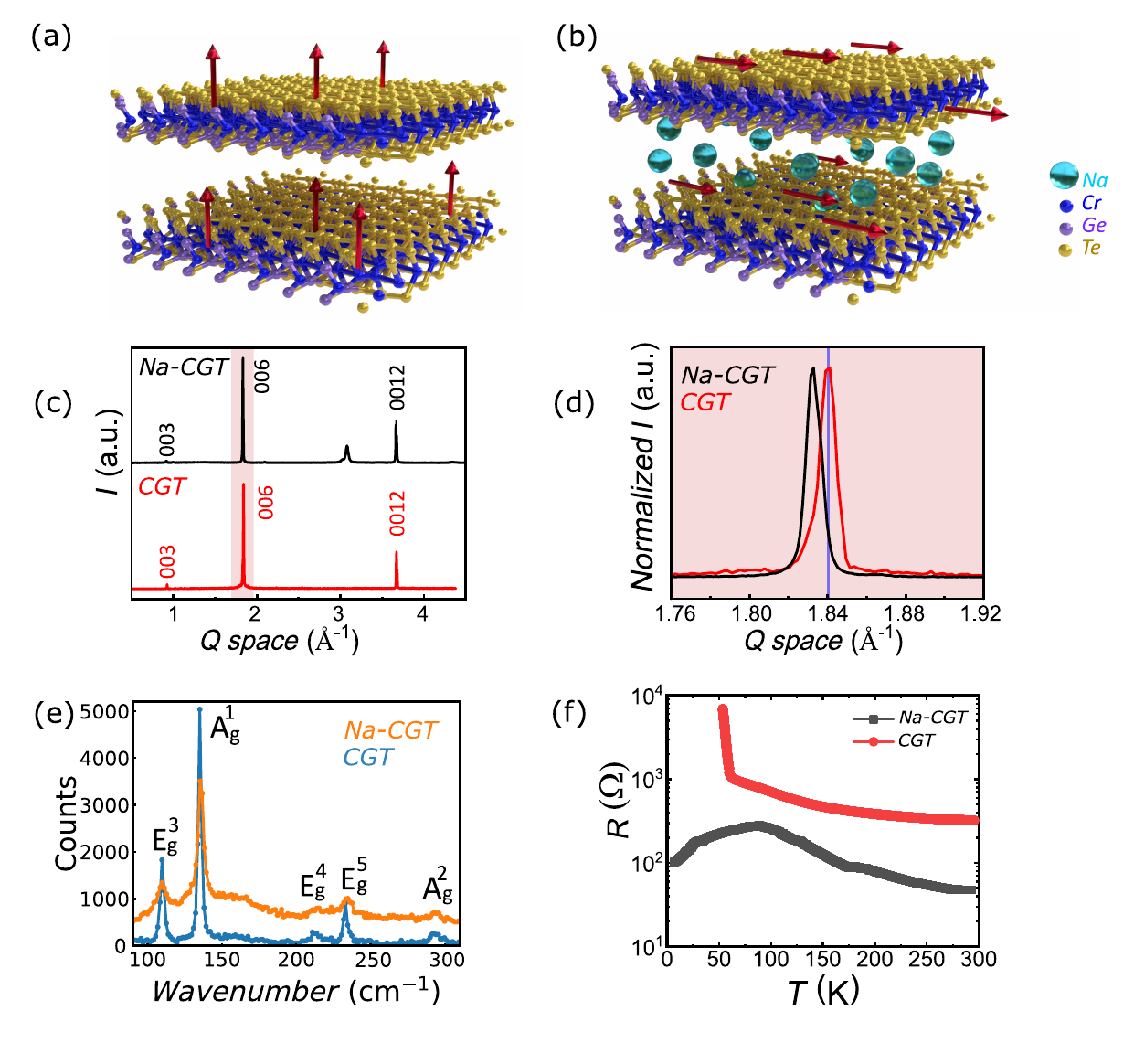}
\caption{(a) Schematic of the pristine CGT with the uniaxial magnetic anisotropy along the out of plane orientation. The red arrows represent magnetic moments along the easy axis. (b) \rev{Illustration} of the Na-intercalated CGT with the magnetic easy plane. (c) XRD of pristine CGT and Na-CGT showing a series of 00l diffraction peaks. The peak around 3 \AA$^{-1}$ in Na-CGT is due to the beryllium sample holder (see Supplementary Information). (d) The magnified section for 006 diffraction peak for comparison between the two materials. (e) Raman spectra for pristine (blue) and Na-intercalated (orange) CGT, showing the major phonon modes for CGT based on assignments from Ref.~\citen{Wang_JACS2019}. (f) Temperature dependence of resistance in CGT and Na-CGT.}
\label{figure1}
\end{figure}

\subsection{Characterisation of Na-Cr$_2$Ge$_2$Te$_6$}

Figure~\ref{figure1} (a $\&$ b) depicts the CGT and Na-CGT systems, respectively, and illustrates the presence of sodium-ions in the vdW gaps. Single CGT crystals were intercalated with sodium ions via the ammonia-based method~\cite{bin2021charge} (details in Methods section). Structural \rev{and chemical} changes in the intercalated CGT are characterised using X-ray diffraction (XRD), Raman spectroscopy and \rev{energy-dispersive X-ray (EDX) spectroscopy (see  Supplementary Information Section 3)}. XRD results in Fig.~\ref{figure1} (c $\&$ d) show that the intercalated sample preserves the original crystalline structure of CGT. The c-axis parameter of pristine CGT is 6.8\,\AA, which corresponds to the distance between the centre of adjacent layers. A downshift is observed for all 00l-peaks after intercalation, indicating an increase in the interlayer spacing by 0.04\,\AA. This small increase in interlayer spacing upon intercalation is also observed in other intercalated compounds~\cite{PaddyCullen2017Ionic}. Raman spectroscopy is also performed for a large number of spatial positions on the sample surface, and the peak structure is maintained, consistent with the indication from XRD. However, quantitative analysis of the three most significant Raman modes ($A^1_g$, $E^2_g$, and $E^5_g$) shows that there are statistically significant changes to phonon energy and lifetime with intercalation. There are no major changes to the three main Raman modes that arise from intralayer vibrations of CGT, as expected, as intercalation should not change the bonding within the layers themselves. The small changes in peak energy and width (which is determined by lifetime) are consistent with carrier doping~\cite{dean2010nonadiabatic,pisana2007breakdown} (see Supplementary Information Fig.S2-S3).

Electrical transport measurements were performed in order to see the effect of carrier doping by the alkali atoms to the system. Figure~\ref{figure1}(f) shows the temperature dependence of resistance for both CGT and Na-CGT. Unlike the standard semiconducting behaviour of temperature dependence in pristine CGT~\cite{zhang2016magnetic,liu2019anisotropic}, Na-CGT displays a semiconductor-like response down to 80\,K, before showing a metallic-like resistance behaviour in the low temperature regime. While complex, this clear change in temperature dependence of resistance further emphasises that electron doping by the intercalation process was successful.  

\begin{figure}[h!]
\centering
\includegraphics[width=18cm]{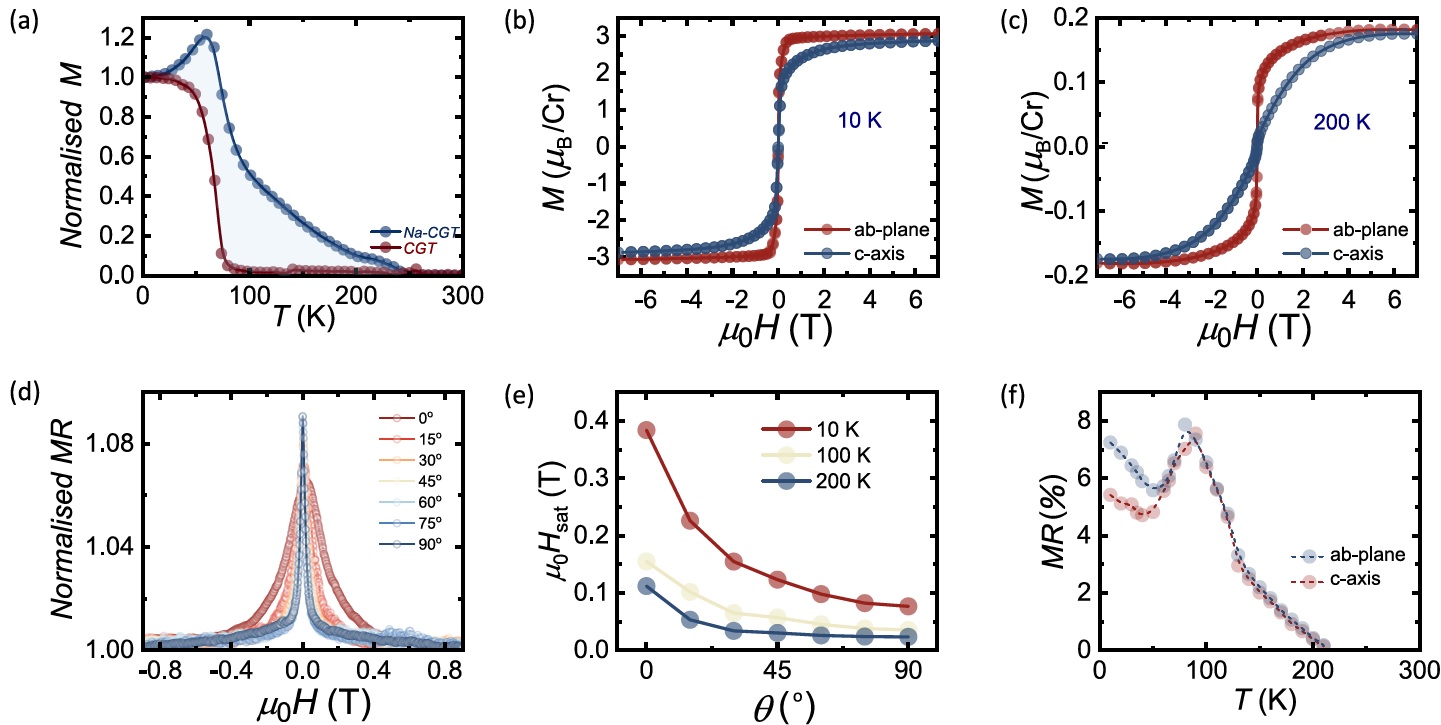}
\caption{(a) Temperature dependence of magnetisation $M$ in CGT and Na-CGT measured during cooling in the absence of a magnetic field. $M$ is normalised at the lowest temperature data point for each scan. (b)\&(c) $M$-$H$ loops of Na-CGT for ab-plane and c-axis directions measured at 10 K and 200 K, respectively. (d) Fixed-angle magnetoresistance (MR) results of Na-CGT for different angle $\theta$ defined from the c-axis direction, for 10\,K. (e) Saturation fields ($\mu_0 H_\text{sat}$) extracted for each MR scan for different $\theta$ and temperature. (f) A plot of the percentage change of MR as a function of temperature in Na-CGT for both ab-plane and c-axis.}
\label{figure2}
\end{figure}

Figure~\ref{figure2} (a) shows temperature-dependent magnetisation along the ab-plane orientation in both CGT and Na-CGT, and demonstrates the impact of Na intercalation on magnetisation in the static regime. In contrast to $T_\text{C}$ of CGT $\sim$ 66\,K~\cite{Khan2019SpinDynamicsPRB}, Na-CGT shows its magnetisation values maintained to a higher temperature up to $\sim$ $240\,K$, implying this as its $T_\text{C}$. \rrev{Our theoretical modeling also reproduces the trend of $T_\text{C}$ enhancement with the introduction of Na intercalated atoms in CGT (see Supplementary Information Fig.S14). Details of the Na distribution between the layers, complexity of the spin interactions and potential electron localisation are factors for such simulations. Experimentally, enhancement of $T_\text{C}$ in CGT} has been reported by organic molecule intercalation (to $\sim$ 208 K)~\cite{Wang_JACS2019} and by pressure (to $\sim$ 250\,K)~\cite{Bhoi_PRL2021} where in all cases metallic transport accompanies the $T_{\text{C}}$ enhancement. Since $T_\text{C}$ scales with the exchange coupling strength, it is possible to interpret in our sample that doped electrons enhance the exchange coupling strength and at the same time these less-localised electrons can contribute to transport properties. In pristine CGT, electrons are fully localised and the super-exchange mechanism that arises from virtual electron excitations between two neighbouring Cr ions via a Te ion with the 90$^\circ$ Kanamori-Goodenough mechanism~\cite{Kanamori_JPCS1959}. Local Hund's rules within both Te and Cr ions stabilise a ferromagnetic order. Through this process, electrons have to be virtually excited into the conduction band via the bandgap, the exchange energy gain by this process is marginal, hence low $T_\text{C}$ in CGT. Doping {\it real} electrons to Cr ions changes their ionic states from Cr$^{3+}$ into Cr$^{2+}$, activating the double-exchange mechanism~\cite{izyumov2001double} that bolsters the exchange interaction. It may also be relevant that since Cr$^{2+}$ is a Jahn-Teller ion, its substitution for Cr$^{3+}$ will lead to local distortions~\cite{nagle2024van} and increased local anisotropy that potentially enhances any spin glass behaviour that we discuss below.  

CGT is known to have uniaxial magnetic anisotropy with the out-of-plane easy axis~\cite{Khan2019SpinDynamicsPRB,zhang2016magnetic}. Magnetisation measurements as a function of magnetic field ($M$-$H$ curves) are performed for Na-CGT and the results at 10 K for the ab-plane and c-axis orientations are shown in Fig.~\ref{figure2} (b). 
%Within the ab-plane, a saturation magnetic moment of 3.06\,$\mu_{\textrm{B}}$/Cr is obtained whereas along the c-axis it is found to be 2.87\,$\mu_{\textrm{B}}$/Cr. 
It is evident that after intercalation, spins prefer to align within the ab-plane compared to the c-axis in CGT, implying the sign change of the uniaxial anisotropy term upon intercalation\rrev{, which is reproduced by our simulations with and without Na intercalants in CGT (see Supplementary Table S5).}   
Figure~\ref{figure2} (c) shows the $M$-$H$ curves of Na-CGT at 200 K, showing a clear magnetic response for both orientations with the sustained uniaxial anisotropy. As a proof of the long-range magnetic order, such a response disappears at 295 K where only a linear $M$-$H$ trend is observed and shown in Fig.S\rev{6}(d) Supplementary Information. 

Electrically conducting behaviour in Na-CGT allows us to probe the magnetic response via magnetoresistance (MR) which was impossible in pristine CGT. We performed angular dependent MR measurements along the sample orientation of in-plane (ab-plane) to the out-of-plane (c-axis) direction, as shown in Fig.~\ref{figure2}(d), showing the clear magnetic anisotropy easy axis (plane) from the c-axis to easy-plane in Na-CGT since the saturation field becomes larger for more out-of-plane measurements. Figure~\ref{figure2}(e) captures this change in magnetic anisotropy for different temperatures (calculation of the saturation field $H_\text{sat}$ is highlighted in Fig.S\rev{12}(b)) of Supplementary Information where $H_\text{sat}$ for c-axis ($0\degree$) increases with decreasing temperature, and so as the difference in $H_\text{sat}$ to the ab-plane ($90\degree$). A sizable change in MR for Na-CGT across a wide temperature range is shown in Fig.~\ref{figure2}(f), which disappears at the temperature where the long range order of $M$ also vanishes in Fig.~\ref{figure2}(a). This measurement reassures consistency in the new intercalated Na-CGT samples and reproducibility of experimental data by means of separate probing techniques.\\

\subsection{Emergent spin-glass nature}
\begin{figure}[b!]
\centering
\includegraphics[width=16cm]{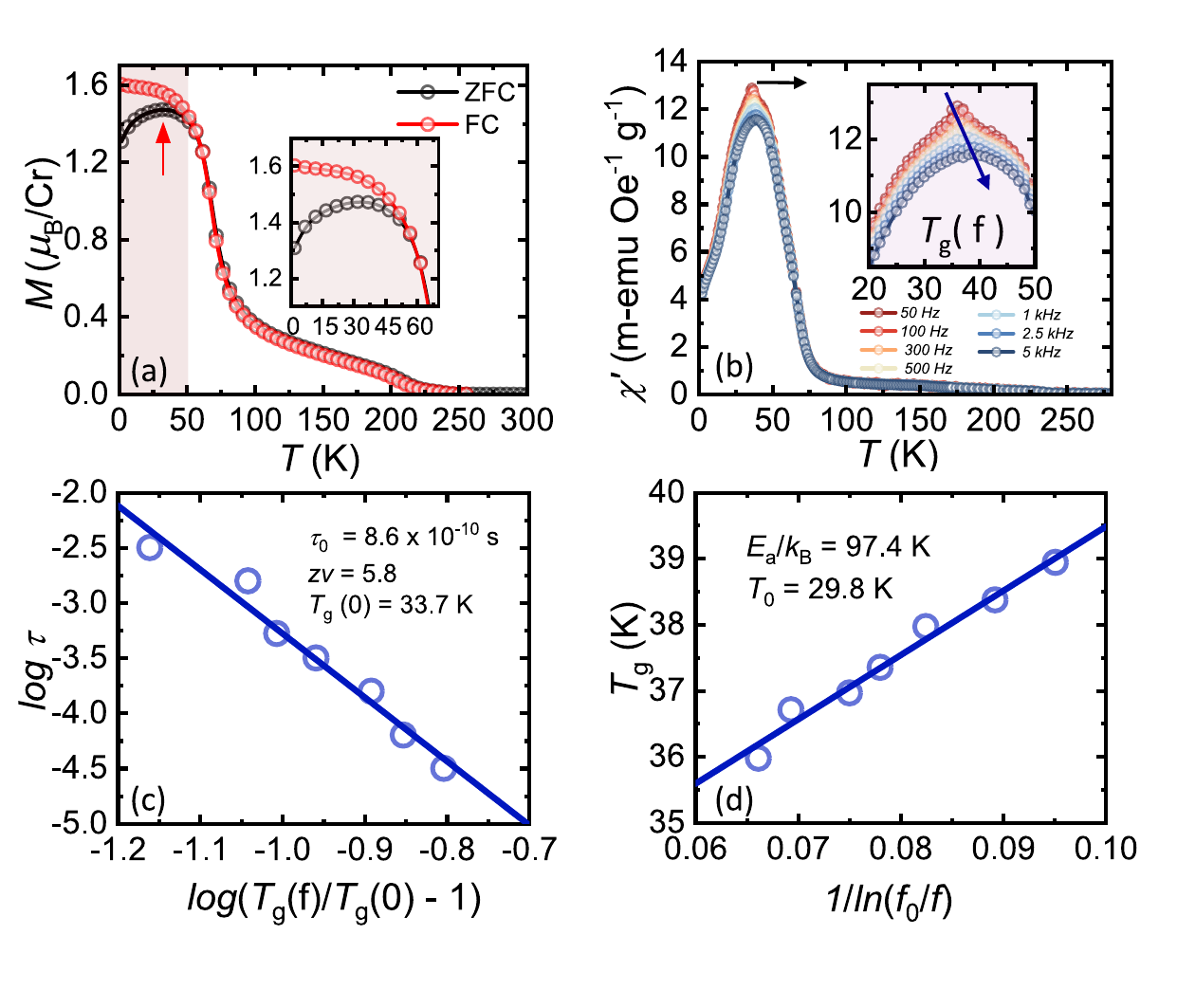}
\caption{(a) $M$ vs $T$ curves in Na-CGT showing a bifurcation feature with different field cooling protocols. (b) Temperature dependence of real part of ac magnetic susceptibility. We define the peak temperature as $T_\text{g}(f)$ that is a function of frequency as shown in the inset. (c) $\&$ (d) Plots of fitting results using $T_\text{g}(f)$ and the function of the scaling power and VF law, respectively. See more details in the main text.}
\label{figure3}
\end{figure}

Another striking property which emerges via the intercalation process is spin-glass like behaviour in Na-CGT, a feature absent in pristine CGT. %It is important to note that spin-glass states have not been experimentally reported in other heavily studied 2D magnetic systems~\cite{hossain2022synthesis,ningrum2020recent}, such as CrI$_{3}$, CrBr$_{3}$, Fe$_5$GeTe$_2$, as well as CGT with carrier doping by organic molecules ~\cite{Wang_JACS2019} and electrostatic gating ~\cite{IvanVerzhbitskiy2020Controlling,Zhuo_AdvMater2021}, and pressure application~\cite{Bhoi_PRL2021,O’Neill_ACSNano2023}.
Spin-glass behaviour in Na-CGT is identified as the divergence of magnetisation in temperature between the zero-field cooled (ZFC) and field-cooled (FC) protocols, at small magnetic fields. As shown in Fig.~\ref{figure3} (a), a bifurcation between the two cases is observed below a temperature at which a steep increase in magnetisation takes places. The magnetisation of the ZFC curve shows a defined broad peak, whereas in the FC protocol, the magnetisation keeps on increasing as the sample is further cooled. The local maximum is observed and we define this as $T_\text{g}$~\cite{bandyopadhyay2006memory,kumar2021cluster}. This bifurcation in the magnetisation is minimised as the size of magnetic field in increased (see Supplementary Information). Such splitting and irreversibility in the magnetic memory of the system are recognised as characteristic features which can arise from various phenomena, including spin-liquid, super-paramagnetic, disordered anti-ferromagnetic and spin or cluster glass states~\cite{Mydosh1993SpinGlassBook,Mydosh2015SpinGlassReviewPaper}. 

In order to gain further insight in this feature potentially representing a spin-glass like state in Na-CGT, ac magnetic susceptibility measurements were performed, where $ \chi_{\text{a.c.}} = \chi^{'} + i \chi^{"} $~\cite{ToppingBlundell2018ACReview}. Figure~\ref{figure3} (b) shows the real (in-phase), $\chi^{'} (T)$, component of the susceptibility, as a function of temperature at different frequencies ranging from 50 Hz to 5 kHz. A distinctive cusp-like anomaly is quite evident in $\chi^{'} (T)$ at low temperatures, which matches to the same temperature region of branching-off observed in Fig.~\ref{figure3} (a). In the inset of Fig.~\ref{figure3} (b), the main feature of $\chi^{'} (T)$ is a small frequency shift in the peak-maximum towards higher temperatures as the frequency is increased. The shift in peak position for a frequency window of Hz-kHz indicates slow dynamics in the system, and this is contrary to ferromagnetic and conventional systems where a similar frequency dependence of $\chi^{'} (T)$ is found only in the MHz regime~\cite{Mydosh1993SpinGlassBook}. Figure~\ref{figure3} (b) also highlights that $\chi^{'} (T)$ for a small magnetic field disappears at high temperatures. The imaginary part (out of phase), $\chi^{"} (T)$, of the ac susceptibility shown in Supplementary Information (Fig.S\rev{9}), is almost 30 times smaller at the peak maximum than $\chi^{'}$. $\chi^{"} (T)$ increases from zero to a finite value near the expected spin-glass like transition and also displays a clear anomaly with an inflection point, which is related to the dissipative nature of the system due to slow magnetic relaxation~\cite{Balanda2013ACReview,Mydosh2015SpinGlassReviewPaper}. 

The observed spin-glass property has formed within the ferromagnetically ordered matrix background. This particular type of spin freezing below a ferromagnetic transition is referred to as the reentrant spin glass state~\cite{Gabay_PRL1981}. Here, the fluctuation of local exchange interaction is sufficient enough to produce the local energy minima that stabilise frozen spin-glass states against the ferromagnetic order~\cite{Verbeek_PRL1978,MURANI_SSC1980}. Phenomenologically, this condition is met when the averaged exchange coupling strength for the $i$th and $j$th site spins <$J_\text{ij}^2$> and its local fluctuation average <$J_\text{ij}$>$^2$ are comparable to each other~\cite{Gabay_PRL1981}, where < · · · > denotes an average over the entire spin system. In Na-CGT, this situation is likely established due to inhomogeneous Na intercalation that induces an electron-doping distribution. Since the exchange coupling strength is a function of carrier density \rrev{in CGT~\cite{IvanVerzhbitskiy2020Controlling,Wang_JACS2019}}, such doping inhomogeneity brings about the fluctuation and hence local competition of the exchange interaction, a key ingredient for the spin-glass states. \rrev{While such atomic scale fluctuation is not directly measured in our experiments, we performed ab-initio simulations to calculate the exchange coupling strengths with Na intercalation, as shown in SI Sec. S7 where we clearly observe the change of the strengths by intercalation. However, as discussed later, the observed spin-glass states are a magnetic cluster type, which is rather difficult to reproduce by atomic-level disorders as we simulated. We speculate that the observed spin-glass states in Na-CGT originate from rich and complex mechanisms, possibly involving other inhomogeneity arising from inhomogeneous doping, such as the spatial distribution of magnetic anisotropy~\cite{IvanVerzhbitskiy2020Controlling,Wang_JACS2019}.  }

For more quantitative analysis \rrev{of experimental results}, a series of analytical procedures is carried out with phenomenological scaling and power-law models~\cite{Mydosh1993SpinGlassBook,Benka2022interplayImp} to further inquest and interpret the data with $T_{\text{g}}$ in $\chi^{'}(T)$ as a function of frequency. The Mydosh parameter~\cite{Mydosh1993SpinGlassBook}, $K$, is measured as the respective variation of $\chi^{'} (T)$ peak-maximum position per frequency, $(\omega = 2 \pi f)$, defined as, $K = \Delta T_{\text{g}}/T_{\text{g}} \Delta\log_{10}(f)$. Here, $\Delta T_{\text{g}}$ corresponds to the difference in experimental temperature values of peak-maximum at high and low frequencies; $\Delta \log_{10}(f)$ is the difference in excitation frequencies for the given $T_{\text{g}}$ value. \textit{K} sets a criterion for differentiating between superparamagetic transition ($K \geq 0.1$) and spin-glass behaviour ($K \leq 0.1$)~\cite{Mydosh1993SpinGlassBook,mulder1982frequency,dormann1999pure, sharma2007magnetic}, and a similar frequency-dependent shift in $T_{\text{g}}$ has not been reported for anti-ferromagnetic or ferromagnetic phases at low $\omega$-values~\cite{Mydosh1993SpinGlassBook}. For Na-CGT, $K$ is experimentally determined to be 0.01 which clearly suggests its spin-glass nature, sitting between the boundary of canonical and cluster spin glasses\cite{bag2018cluster,bag2020studyCluster}. 

In the next step, the standard theory of dynamic scaling near the phase transition temperature $T_{\text{g}} (f)$ for given excitation frequencies is employed~\cite{Mydosh1993SpinGlassBook,Benka2022interplayImp}. This theory relies on a power law dependence of relaxation time (experimental attempt time), $\tau = 1/2\pi f$, as\rrev{: 
\begin{equation}
\tau = \tau_{0} [T_{\text{g}}(f)/T_{\text{g}}(0) - 1]^{zv},
\label{eq:power}
\end{equation} 
}where $\tau_{0}$ is the characteristic relaxation time related to the microscopic flipping time for dynamical fluctuating entities, such as a single moment or a cluster. $T_{\text{g}} (0)$ corresponds to the freezing temperature in the limit of $f \rightarrow 0$, and $zv$ is the critical exponent. The power law equation is simplified into a linear form using a logarithmic base, and $T_{\text{g}} (0)$ is adjusted in order to get the best linearity~\cite{Benka2022interplayImp}, where the goodness of fit is determined from the highest $R^{2}$-coefficient value. Figure~\ref{figure3} (c) shows the result of this fitting process of the power law. $T_{\text{g}} (0)$ is found to be 33.7\,K, and then using this value, a linear least-squares fitting method is utilised to extrapolate $\tau_{0} = 8.6 \times 10^{-10}$\,s and $zv = 5.8$. The obtained value of $\tau_{0}$ for Na-CGT lies in the range ($10^{-7}$ to $10^{-11}$\,s) typical for cluster spin-glass systems~\cite{bag2020studyCluster,malinowski2011spin} compared to canonical spin glasses~\cite{wang2004spin,vijayanandhini2009spin} ($10^{-12}$ to $10^{-14}$\,s) where the fluctuation of individuals moments happens at a faster rate. This indicates slower spin dynamics in Na-CGT at the transition temperature and it is most likely due to the presence of interacting spin-clusters in the system instead of individual spins. The value of \textit{zv} for typical systems with spin-glass behaviour is reported between 4 and 12~\cite{bag2020studyCluster,vijayanandhini2009spin}, and indeed this is also the case for Na-CGT, providing another experimental evidence for the cluster spin-glass nature in Na-CGT.
\begin{figure}[b!]
\centering
\includegraphics[width=16cm]{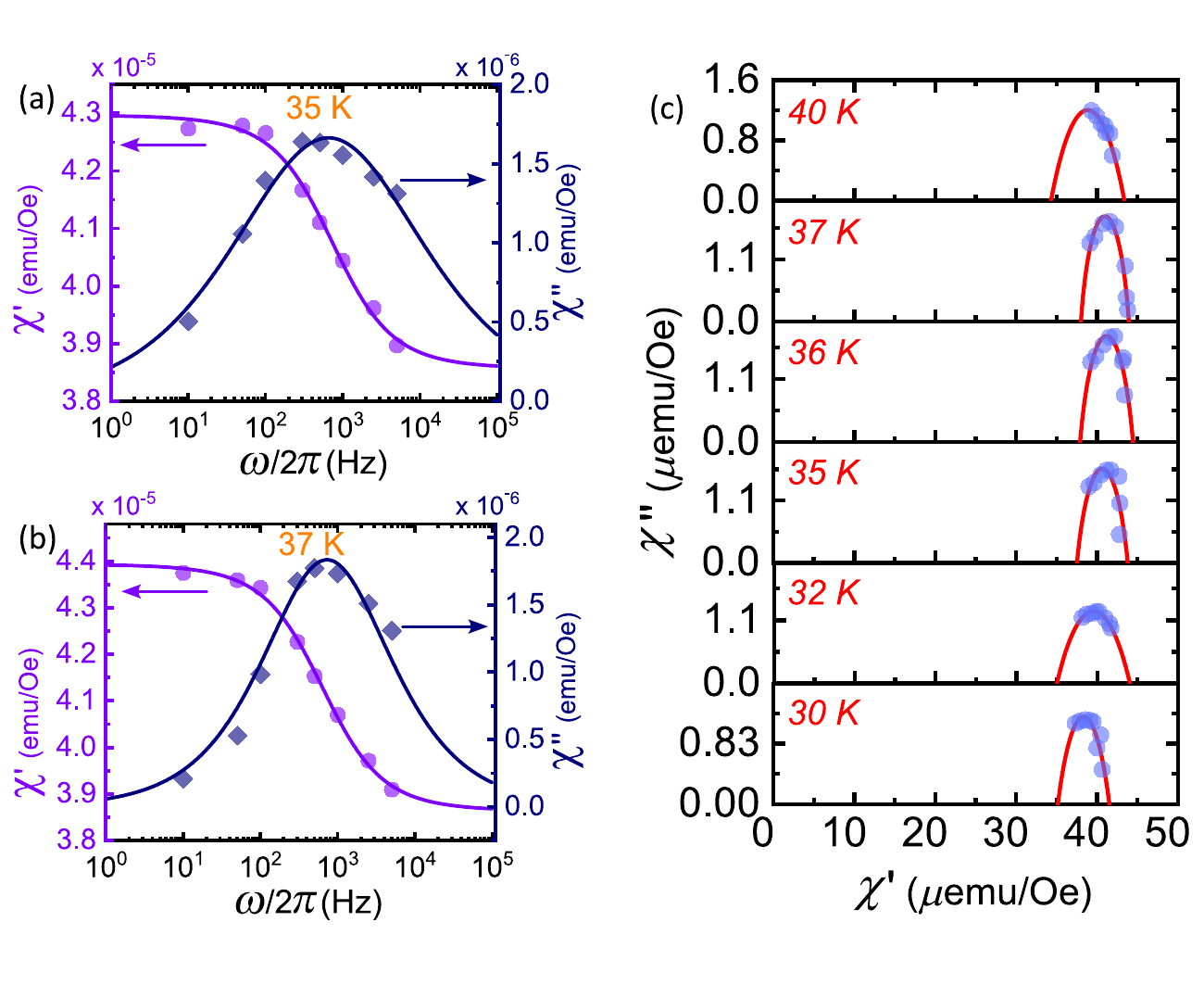}
\caption{(a) $\&$ (b) Plots of both the real and imaginary parts of ac susceptibility ($\chi'$ and $\chi"$) as a function of frequency for 35\,K and 37\,K, respectively. Curves using the best fit parameters calculated by Eqs. S3 and S4 are shown. (c) $\chi"$-$\chi'$ plots for different temperatures together with the fitted curves generated by the Cole-Cole equation (Eq.(\ref{Eq:1})).}
\label{figure4}
\end{figure} 

For magnetically interacting clusters, a modified form of the Arrhenius law, known as the Vogel-Fulcher(VF) law~\cite{Balanda2013ACReview,Benka2022interplayImp} is used instead, to analyse ac susceptibility results. The Arrhenius law describes the relaxation of non-interacting spins~\cite{Mydosh1993SpinGlassBook,Balanda2013ACReview}, and it is highlighted in Supplementary Information (Fig.S\rev{10}) that even though the data seem to fit an Arrhenius law, the fit would imply unphysical values for the relaxation time and activation energy, suggesting that a description in terms of non-interacting spins is inadequate. The phenomenological VF law for the excitation frequency, $f$, takes the following form, $f = f_{0} \hspace{0.2cm} \textrm{exp}[-E_{\text{a}}/k_{\text{B}}(T_{\text{g}}(f)-T_{0})]$, where the characteristic frequency, $f_{0} = 1/2\pi \tau_{0}$, is obtained using the characteristic relaxation time from the scaling law, with $E_{\text{a}}$ and $k_{\text{B}}$ being the activation energy and the Boltzmann constant respectively. The constant, $T_{0}$, is known as the VF temperature and attributed to inter-cluster interaction strength, and it is constrained having a value between 0\,K to $T_{\text{g}}$~\cite{kumar2014evidence}. Figure~\ref{figure3} (d) shows a linear fit to the experimental data, where the parameters obtained from the least-squares fitting from the slope and intercept yield $E_{\text{a}}/k_{\text{B}} = 97.4$\,K and $T_{0} = 29.8$\,K, respectively. The agreement of the VF law to the experimental data and a finite value of $T_{0}$ insinuates interaction among the clusters, as the ratio between $T_{0}$ and $E_{\text{a}}/k_{\text{B}}$ provides information about the interacting strength between the clusters~\cite{Balanda2013ACReview,Benka2022interplayImp}. In the case of Na-CGT, $T_{0} \sim 0.3 \hspace{0.2cm} E_{\text{a}}/k_{\text{B}}$, which lies in the intermediate regime therefore highlighting a finite interaction between the dynamic entities.

\rrev{Furthermore}, the Casimir-du Pr\'e equations are utilised ~\cite{Mydosh1993SpinGlassBook,ToppingBlundell2018ACReview}, in order to describe the dependence of both real and imaginary parts of the ac susceptibility on the excitation frequency at a given temperature. Figures~\ref{figure4} (a) and (b) show good agreement between the experimental data for both real/imaginary parts and the modified equations (Equations S3 $\&$ S4) for different temperatures. The slow dynamics, arising in the vicinity of 35\,K, are described by the generalised Debye model that takes into account the entities (i.e. spin-clusters for Na-CGT) having finite interactions leading to a system having a distribution of relaxation times due to the presence of cooperative effects~\cite{ToppingBlundell2018ACReview}. $\chi"$ is plotted against $\chi'$ at a single temperature value and for different frequencies as shown in Fig.~\ref{figure4} (c), and we fit them by the following Cole-Cole equation:   

\begin{equation}\label{Eq:1}
\chi^{''}(\chi^{'}) = - \frac{(\chi_{0} - \chi_S)}{2\tan[(1-\alpha)(\pi/2)]} + \sqrt{(\chi^{'} - \chi_{S})(\chi_{0}-\chi^{'}) + \frac{(\chi_{0}-\chi^{'})^{2}}{4(\tan[(1-\alpha)(\pi/2)])^{2}}},
\end{equation}
where $\chi_{0}$ represents the susceptibility in the limit of zero frequency, known as isothermal susceptibility; $\chi_{\text{S}}$ is the adiabatic susceptibility for the high frequency limit; $\alpha$ is a dimensionless parameter to represent its relaxation nature with 0 being the case for single relaxation time and 1 being for a broad distribution of relaxation times. The fact that the fit curves show an arc centered away from the origin indicates the presence of a distribution of relaxation times, compared to a semi-circle appearing for the system with a single relaxation time~\cite{ToppingBlundell2018ACReview}. The presence of such distribution can be attributed to magnetic clusters with different sizes and anisotropies, each experiencing different relaxation times in the whole system. Further quantitative analysis are available in Supplementary Information. This series of analysis confirms that at low temperatures (in the vicinity of $\sim$ 35\,K), intercalation of CGT by Na-atoms leads to the co-existence of a cluster spin-glass transition with a background ferromagnetic state. The analysis point towards a progressive freezing of spin clusters and slower dynamics.

\rrev{Finally, we compare the spin-glass behaviour observed in Na-CGT and those in other vdW material systems reported so far~\cite{GOOSSENS_JMMM2013,Takano_JAP2003,MASUBUCHI_JAC2008,Graham_PRMater2020,McGuire_PRMater2018,wang2023spin}. A critical difference between ours from the other vdW spin-glass systems is on the way to generate magnetic frustration. While our study demonstrates intercalation of non-magnetic atoms as a new route of generating magnetic frustration, other existing studies have used substitution of magnetic elements as a common method for creating spin-glass states. An example is 3d-transition metal phosphorus trisulfide XPS$_3$ where X is substituted between two elements (such as Fe$_{0.5}$Ni$_{0.5}$~\cite{GOOSSENS_JMMM2013} and Mn$_{0.5}$Fe$_{0.5}$~\cite{Takano_JAP2003,MASUBUCHI_JAC2008,Graham_PRMater2020}). There is a recent study by Wang et al.~\cite{wang2023spin} where amorphous Cr-Ge-Te hosts spin glass behaviour and its driving mechanism is due to the presence of structural motifs corresponding to the known crystalline phases Cr-Te$_\text{n}$ (n = 1, 2 and 3) with different Cr-Te-Cr bonding angles having a series of exchange interaction strengths. The importance of our study is that we maintain the original crystalline structure during the intercalation (as evident by XRD results) and generate spin-glass states without changing the Cr-Te-Cr bonding angles. It is worth highlighting that the relaxation rate extracted by the power law analysis (Eq.(\ref{eq:power})) in the amorphous Cr-Ge-Te system is $\tau_0=8.9\times10^{-11}$s which is one order faster than ours ($8.6 \times 10^{-10}$\,s) based on the same analysis. This difference in relaxation rates is consistent with our classification of the types of spin-glass state in amorphous Cr-Ge-Te (short-lived canonical spin-glass, with a small moment and rapid relaxation rate) and Na-CGT (cluster spin-glass with large moment and slower relaxation).}

\section{Conclusion}
In summary, we show the unique feature of Na intercalation processes to significantly modify the transport/magnetic properties of vdW magnetic material CGT. We discovered that the intercalated Na-CGT hosts emerging magnetic states, i.e. the spin-glass states that are absent in pristine CGT. Such a spin-glass phase is a reentrant spin-glass system that coexists with the ferromagnetically ordered background. Dynamic susceptibility measurements analysed by multiple models all suggest the nature of cluster spin-glasses where spins are interacting to each other to behave collectively with the distribution of relaxation timescales. This study sheds light on the unique feature of intercalation to generate magnetically frustrated systems by inhomogeneous carrier doping. \rev{The ammonia-solution-based, alkali-atom intercalation techniques presented in this study can be performed with other alkaline elements such as K and Ca, for further exploring the tunability of magnetic ground states and induced frustration with different intercalants.} \rrev{We envisage that unique spin-glass properties generated by this alkali-atom intercalation route will be actively discussed and identified once a good number of experimental demonstrations have been reported.} This new way of tuning vdW materials might be effective beyond the field of magnetism, where inhomogeneity will play an active role for determining material ground states.

\section{Methods}
Direct vapor transport method was used to synthesise the single-crystalline flakes of Cr$_2$Ge$_2$Te$_6$. Pre-mixing with a molar ratio of $20:27:153$ of high-purity elemental Cr ($99.9999\%$ in chips), Ge ($99.9999\%$ in crystals), and Te ($99.9999\%$ in beads) was carried out and then sealed in a quartz ampule at high vacuum ($\sim 10^{-5}$ Torr). The ampule was loaded in a single-zone furnace with increasing the temperature at a rate of 2 K/min reaching up to 1273 K. The furnace was set to slow cooling with the rate of 5 K/hour down to 673 K in order to obtain a low defect crystal growth. Then, nature cooling process was used with switching off the furnace. From the flux buildup, the single crystal flakes were separated and then stored in an inert atmosphere in a glovebox.\\ 

\rev{We used established low-temperature intercalation techniques~\cite{Ding_AdvMater2001,PaddyCullen2017Ionic} to synthesise Na-CGT. Macroscopic CGT crystals were outgassed under vacuum to a pressure less than 10$^{-6}$ mbar at 100$^\circ$C overnight, in a custom-made air-tight glass tube. In an Ar glovebox, the mass of outgassed crystal was measured, followed by adding Na into the glass tube at a molar ratio of Na:CGT = 1:1. To be specific, we used mass ratio of Na:CGT = 4.7 mg : 208 mg and another batch of 4.3 mg: 184.3 mg. The glass tube was then submerged in a bath at  approximately -50$^\circ$C, where liquid ammonia was condensed onto CGT and Na. The solution was maintained for a period of approximately 72 hours to allow the intercalation to fully complete to form a crystal with the atomic ratio of Na$_1$Cr$_2$Ge$_2$Te$_6$. We can confirm the completion of intercalation by the color change of the ammonia solution~\cite{PaddyCullen2017Ionic,Zurek_Ange2009}. During intercalation, the alkali-metal-ammonia solutions display a deep blue color, arising from the solvated electrons. As the reaction progresses the blue colour is gradually lost as the electrons move onto the layers driving concomitant intercalation. Eventually we observed an orange solution, which can be attributed to the presence of sodium polytelluride excess byproduct. Thus, we can confirm that the intercalation of sodium into the CGT matrix has occurred. Subsequently, the ammonia was slowly extracted via cryo-pumping until intercalated Na-CGT crystals were left.}\\

X-ray diffraction was performed using a Philips X'pert MPD diffractometer with Cu K$_\alpha$ radiation ($\lambda=1.5406$\,\AA{}) in reflection geometry. For the measurement on air-sensitive Na-CGT, the sample was loaded into a berylium dome holder with O-ring type seal inside an Ar glovebox for measurement. Raman spectroscopy was performed using a Renishaw In-Via microscope equipped with a 488\,nm laser and a 2400 I/mm grating, with a x20 magnification. Samples were measured in an air-free glass sample holder, loaded in a glovebox.\\

A Quantum Design Magnetic Properties Measurement System (MPMS-3) was used for dc variable-temperature susceptibility and isothermal magnetisation measurements. Measurements were taken during warming after cooling in zero field (ZFC-W) and in the measuring field (FC-W), where the measuring field was 200\,Oe. Isothermal magnetisation measurements were performed in an external magnetic field between $\pm$ 70\,kOe. A Quantum Design 9 T Physical Properties Measurement System (PPMS) using the ACMS-II option was used for ac susceptibility. Measurements were taken in a static base field with an oscillating ac driving field, where we used a combination of either 10\,Oe and 10\,Oe, respectively, or 50\,Oe and 15\,Oe, respectively. ac measurements were performed with the driving field oscillating at frequencies of 50, 100, 300, 500, 1000, 2500, and 5000\,Hz. \\

For dc susceptibility measurements where external field is applied perpendicular to the layers of the crystal. The crystal was held tightly in place between two pieces of cling-film within a plastic cap which was suspended in the centre of a straw. For the inplane dc susceptibility measurements, the crystal was attached to the flat longitudinal edge of a semi-cylindrical quartz rod by Kapton. For ac susceptibility measurements, the magnetic field was not applied along a particular axis. \\

The electrical transport and magnetotransport measurements are performed by the four-point measurement technique ~\cite{firebaugh1998investigation,dulal2019weak}, as to characterise the electronic properties of the sample. A Keithley 2450 source meter is used to apply the current to the outer most connections and contact free resistance is measured by the two inner electrical connections. A copper sample box is used, inside which the sample is mounted. The sample stage is then placed on a probe copper arm inside of a closed cycle helium cryostat (Oxford Instruments), with a variable temperature range of 5 - 300\,K. The probe arm of the cryostat sits between the poles of an electromagnet, which can produce a homogeneous magnetic field in the sample region. \\ 

\rrev{Theoretical modeling of the exchange interactions before and after the Na-intercalation of CGT was undertaken to elucidate the main factors inducing the appearance of competing spin interactions (e.g., isotropic and anisotropic exchange, anisotropy energies, Dzyaloshinskii-Moriya interaction - DMI) responsible for the formation of the spin-glass state, enhancement of $T_{\text{C}}$, and change of magnetic axis. See Supplementary Information Sec S7 for details.   }

\section*{Acknowledgements}
S.K., C.A.H. and H.K. acknowledge support from the UK Engineering and Physical Sciences Research Council (EPSRC) via EP/T006749/1 and EP/V035630/1.
E.S.Y.A. would like to thank the EPSRC for studentship funding from the Centre for Doctoral Training in Advanced Materials Characterisation (EP/S023259/1).
L.A.V.N-C acknowledges a scholarship EP/R513180/1 to pursue doctoral research from EPSRC. A.S. thanks  JSPS Postdoctoral fellowship for research in Japan (P21777).
Magnetic measurements at the University of Cambridge were made on the EPSRC Advanced Characterization Suite (funded under EP/M0005/24/1).
E.J.G.S. acknowledges computational resources through CIRRUS Tier-2 HPC 
Service (ec131 Cirrus Project) at EPCC (http://www.cirrus.ac.uk) funded 
by the University of Edinburgh and EPSRC (EP/P020267/1). EJGS acknowledges the EPSRC Open Fellowship (EP/T021578/1), and DIPC for funding support. 
A.G.L. acknowledges financial support from the Spanish Ministry of Science, Innovation and Universities under grant No. PRX19/452.

\section*{Author contributions statement}
S.K. and H.K. conceived the idea of the project. I.V. and G.E. synthesised and provided pristine CGT crystals. E.S.Y.A. and C.A.H. intercalated Na-ions in the CGT crystals. E.S.Y.A. performed Raman and XRD experiments. 
\rrev{E.J.G.S. and S.G. provided theoretical modeling and contributed to the explanation of the spin-glass state.}
\rev{E.S.Y.A. and L.A.V.N-C. conducted EDX characterisation and analysis.} L.A.V.N-C. conducted dc-magnetometry and ac susceptibility experiments with supervision of S.E.D. S.K. performed the electrical transport and magnetotransport experiments. Data analysis and interpretations were carried out by S.K., E.S.Y.A., L.A.V.N-C., H.K. and all other co-authors.  S.K., E.S.Y.A., L.A.V.N-C., A.S. and H.K. wrote the manuscript with input from the other co-authors.

%\bibliography{sample}

\newpage
\begin{center}
\newcommand{\todo}[1]{\textcolor{orange}{#1}}

\textbf{\LARGE Supplementary Information for ``Spin-glass states generated in \rrev{a} van der Waals magnet by alkali-ion intercalation"}
\end{center}

\setcounter{figure}{0}
\setcounter{section}{0}
\setcounter{equation}{0}

\renewcommand{\theequation}{S\arabic{equation}}
\renewcommand{\thefigure}{S\arabic{figure}}
\renewcommand{\thetable}{S\arabic{table}}
\renewcommand{\thesection}{S\arabic{section}}
\let\oldcite\cite
\renewcommand*\cite[1]{\textsuperscript{\textcolor{blue}S}
{\oldcite{#1}}}

\section{X-ray Diffraction}

\begin{figure}[ht!]
\centering
\includegraphics[width=18cm]{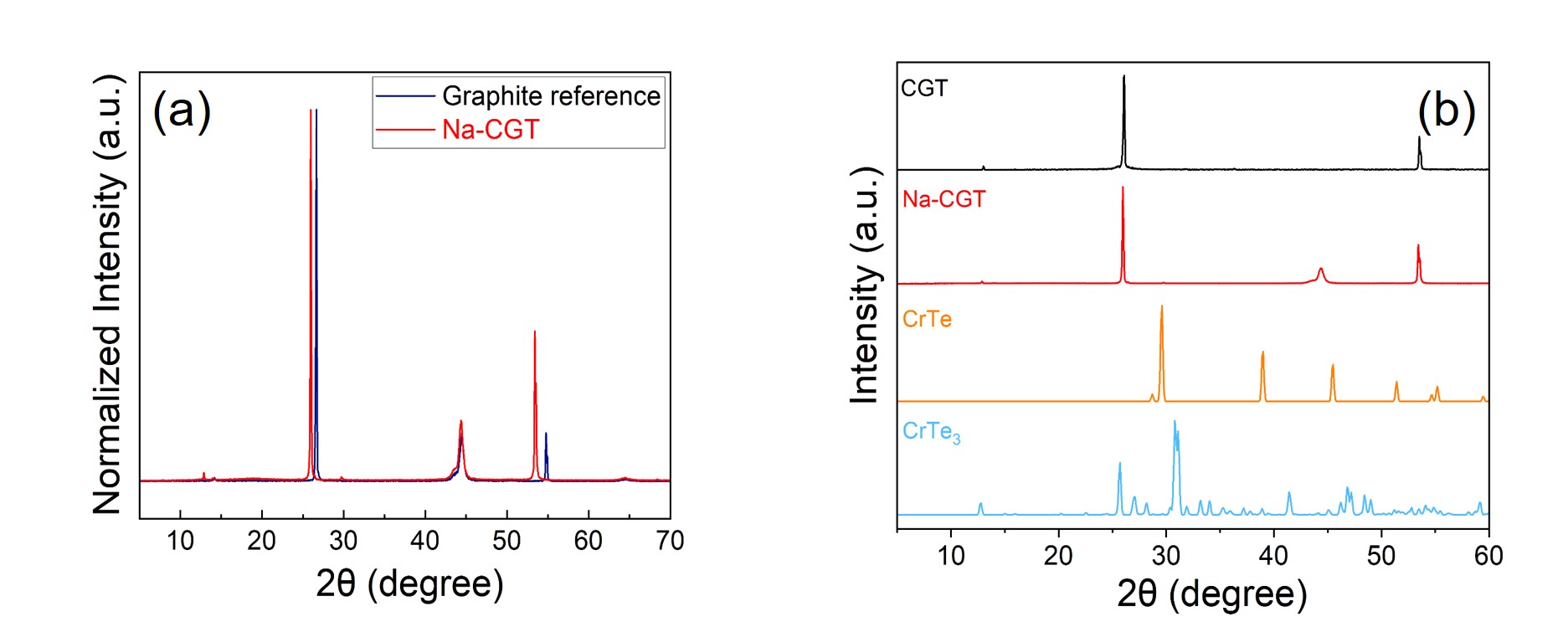}
\caption{X-ray diffraction peaks: (a) XRD patterns of Na-CGT (red) and graphite (black), both performed inside a beryllium dome sample holder. The graphite reference shows that the peak at 44$ ^{\circ}$ in both diffraction patterns can be attributed to the holder~\cite{stefaniak2004characterization,ain2019systemic}.(b) Experimental XRD patterns of CGT, Na-CGT, compared with calculated patterns for CrTe~\cite{nagasaki1969pressure}, CrTe$_3$~\cite{engels1982neue}. This comparison rules out the presence of Cr$_{\text{x}}$Te$_{\text{y}}$ compounds in the intercalated CGT.  }
\label{XRD_Supp_Fig1}
\end{figure}

In order to distinguish background peaks from CGT 00l diffraction patterns, a graphite reference was also measured in a beryllium dome. The peaks shown in Figure~\ref{XRD_Supp_Fig1} (a) corresponds to graphite peaks from literature, with the exception of the peak at 44$ ^{\circ}$ which is present in both diffraction patterns. Hence, the peak at 44$ ^{\circ}$ can be attributed to the sample environment, as opposed to Na-CGT. 

To rule out presence of some common impurities in pristine and intercalated CGT, calculated XRD patterns with Cr-Te compounds were compared with the experimental data. As shown in Figure~\ref{XRD_Supp_Fig1} (b), there is no overlapping of significant peaks between CGT and Cr-Te compounds.  

\section{Raman spectroscopy analysis}

The analysis of the Raman spectroscopy data was performed using the \textsc{Python 3}~\cite{10.5555/1593511} package \textsc{SciPy}~\cite{2020SciPy-NMeth}. Using the \verb|curve_fit| function, a Lorentzian peak-shape was fitted to each of the modes studied, defined as:

\begin{equation}
   I(x) = \frac{A}{\pi} \frac{\Gamma/2}{(x-x_0)^2+(\Gamma/2)^2}
\end{equation}

where $x_0$ is the centre of the peak, $\Gamma$ is the full width at half maximum (FWHM), and $A$ is proportional to peak height.

\begin{table}[h]
	\begin{tabular}{c c c c c c c}
			\toprule
			 & \multicolumn{3}{|c|}{Centre (cm$^{-1}$)} & \multicolumn{3}{|c|}{FWHM (cm$^{-1}$)}    \\ 
     Material & $A_g^1$ & $E_g^2$ & $E_g^5$  & $A_g^1$ & $E_g^2$ & $E_g^5$  \\ 
			\midrule
CGT & 135.17(11) & 109.85(8) & 233.09(18) & 2.6(2) & 3.5(6) & 3.4(7)  \\			
   Na-CGT & 136.03(16) & 109.8(6) & 234.2(5) & 3.8(4) & 4.4(11) &  7.1(13) \\
			\bottomrule
		\end{tabular}
		\caption{\label{raman_SI_table}The average peak position and full width half maximum (FWHM) for the $A_g^1$, $E_g^2$, and $E_g^5$ modes for pristine and Na-intercalated CGT, with standard deviation as error presented in brackets to the same number of decimal places as the value. 25 and 121 spectra were used to calculate the values for pristine and Na-intercalated CGT, respectively.}
\end{table}

\begin{figure}[h]
\centering
\includegraphics[scale=0.85]{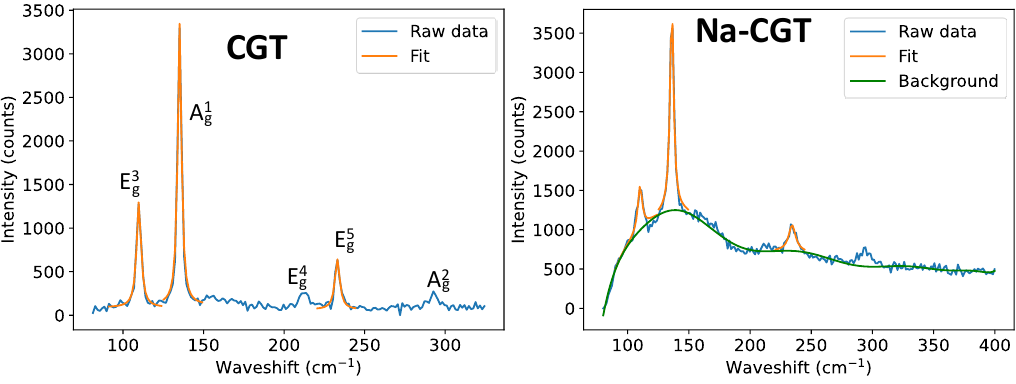}
\caption{Plots showing the fits for CGT and Na-CGT.}
\label{SI_raman1}
\end{figure}

Raman spectra from map scans of the two samples, with spatial resolution 10\,$\mu$m, were studied quantitatively in this way. A fit example is shown in Figure~\ref{SI_raman1}. The spectra for Na-intercalated CGT exhibited significant background throughout the measured range, which was not present in the spectra for the pristine sample. To account for this, a 12-order polynomial was fit to the regions of the data where there were no peaks. This background was subtracted before peak fitting occurred. 

\begin{figure}[p]
\centering
\includegraphics[scale=0.85]{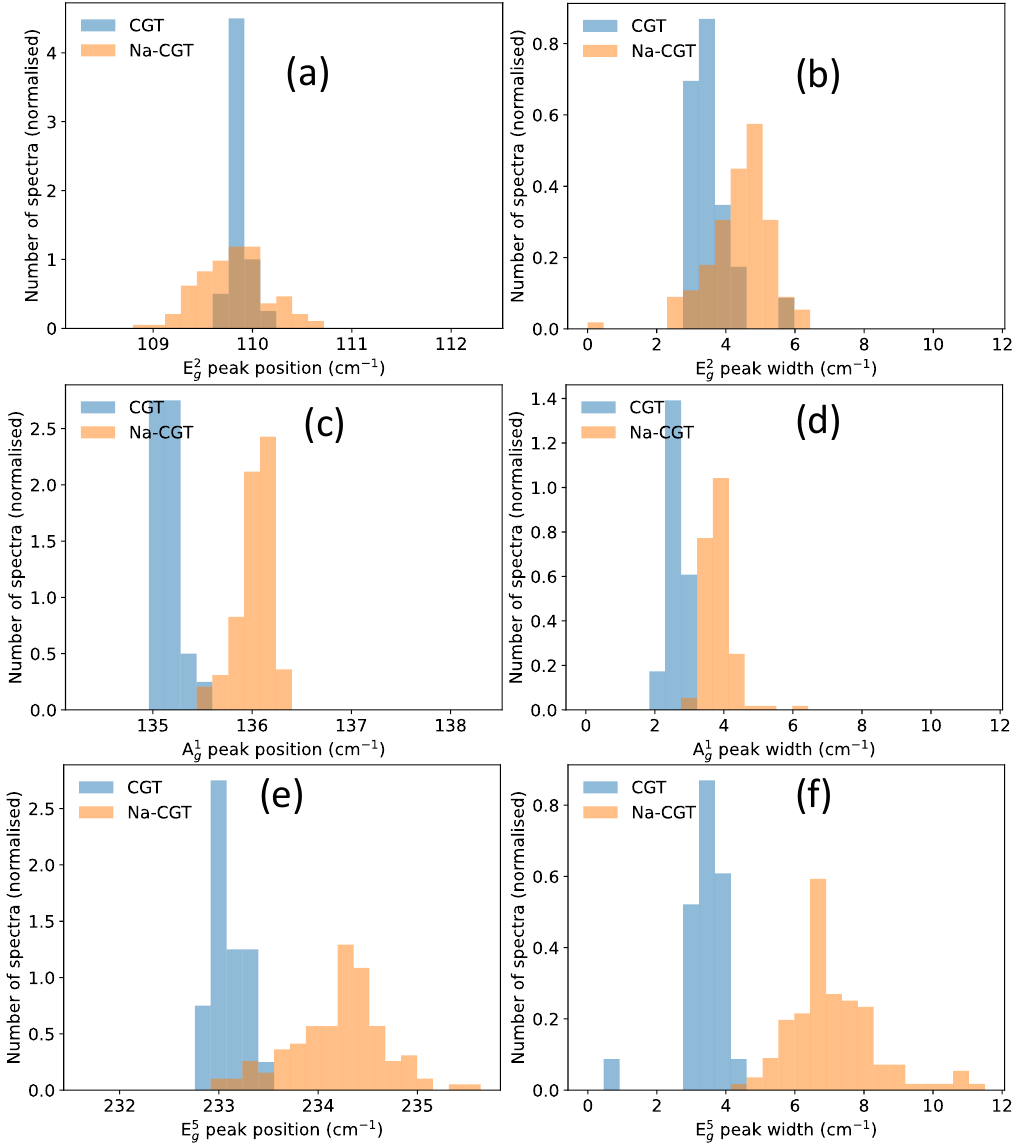}
\caption{Histograms showing the peak position (a,c,e) and peak FWHM (b,d,f) of the $A_g^1$, $E_g^2$, and $E_g^5$ Raman modes.}
\label{SI_raman2}
\end{figure}

For each spectrum in the map scans, the fit was performed to obtain $x_0$ and FWHM for each of the $A^1_g$, $E^2_g$, and $E^5_g$ peaks. This was used to calculate an average and standard deviation, which is tabulated in Table~\ref{raman_SI_table}. The full distributions are shown in Figure~\ref{SI_raman2}. 

From this table it can be seen that there is a statistically significant increase in peak position for the $A^1_g$ and $E^5_g$ modes; within error, the $E^2_g$ mode does not change with intercalation. This consistent, statistically-significant change demonstrates that intercalation has taken place. However, it is a surprising result. With intercalation of an organic molecule in CGT, the Raman peaks decreased in energy due to the addition of electrons to the anti-bonding orbitals which weakened the bonds~\cite{Wang_JACS2019_S}; this is also common to electron-doped layered materials, for instance gated~\cite{pisana2007breakdown_S} or intercalated graphite~\cite{dean2010nonadiabatic_S}. We speculate that this may occur to increased thermal conductivity of Na-intercalated CGT, leading to better dissipation of the heat from the laser.

We also observe an increase in peak-width with intercalation. This is similar to the result for intercalated graphite~\cite{dean2010nonadiabatic_S}. It likely results from the reduced lifetime of the phonon due to electron-phonon coupling, though an alternative explanation could be spatial inhomogeneity within the sample. We note that the degree of the peak broadening depends on the different modes, implying its complex origins including spatial disorder/defects and electron-phonon coupling. The intercalation of an organic tetrabutyl ammonium (TBA) molecule into CGT~\cite{Wang_JACS2019_S} led to the emergence of a new phonon mode while no new mode emerges in our Na intercalated CGT. We attribute this difference that TBA has a Raman-active mode associated with its intra-molecular vibrations, whereas Na is a single atom.

\clearpage

\section{\rev{Energy dispersive X-ray analysis}}

\rev{Electron dispersive x-ray spectroscopy (EDX) was performed to investigate the local stoichiometry of Na-intercalated CGT. A Tescan MIRA3 SEM was used for the measurements. To avoid air exposure and subsequent sample degradation, each Na-intercalated CGT sample was loaded into an airtight sample-holder and transported to the SEM. It was not necessary to coat samples with conductive material and an accelerating voltage of 10 kV was used.}

\begin{figure}[b!]
\centering
\includegraphics[width=15.2cm]{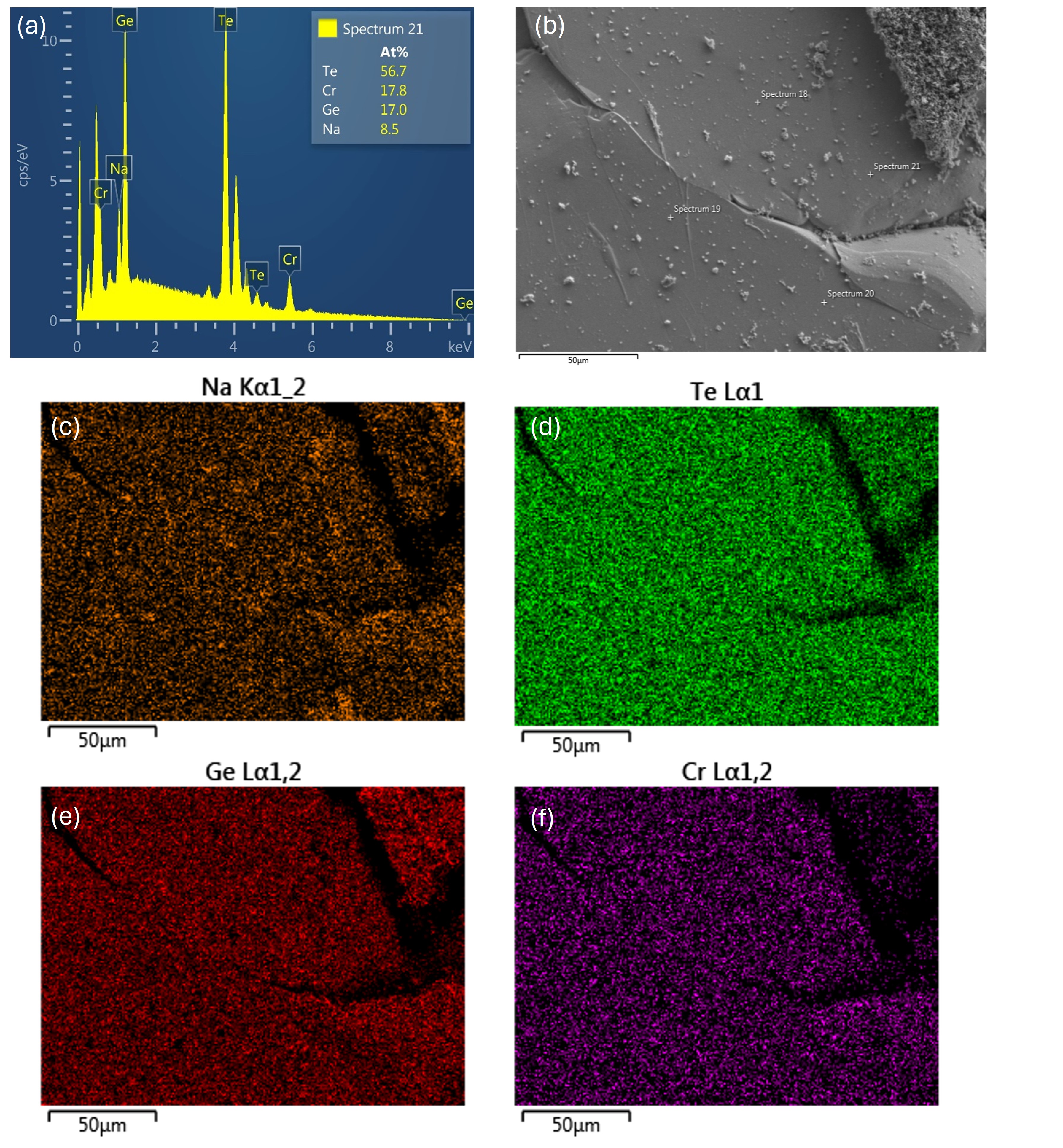}
\caption{\rev{(a) An example EDX spectrum with Na (8.5at\%), Cr (17.8at\%), Ge (17.0at\%), and Te (56.7at\%) contributions. (b) An SEM image of the crystallite surface, with Spectrum 21 corresponding to the point of the spectrum in Fig. S4(a). The distribution of x-ray intensity corresponding to signatures of (c) Na, (d) Cr, (e) Ge, and (f) Te, at the same point shown in Fig. S4(b).}}
\label{edxfig1}
\end{figure}

\rev{As discussed in Method section in the main manuscript, the expected stoichiometric ratio was 1:2:2:6 for Na:Cr:Ge:Te, from the ratio of source materials used in the Na intercalation. Figure~\ref{edxfig1}(a) shows a representative EDX spectrum at point 21 in the SEM image in Fig.~\ref{edxfig1}(b), which shows the presence of Na (8.5at\%), Cr (17.8at\%), Ge (17.0at\%), and Te (56.7at\%), which conforms successful Na intercalation that produces a similar composition ratio to the expected stoichiometry. It is worth noting that the Na peak is relatively weak, impacting on the accuracy of the Na atomic percentage and this ratio. The inherent lack of accuracy of SEM-EDX for elemental analysis is also evident in the Cr:Ge ratio, although such variation is common in SEM-EDX measurements due to a range of factors~\cite{RN169}. Element-sensitive mapping of the Na-CGT as shown in Fig.~\ref{edxfig1}(c)-(f) show each element is distributed across the sample, with intensity variation typical for SEM-EDX maps, arising from uneven surface topography~\cite{RN169}. }

\section{DC-SQUID Measurements} 
\begin{figure}[ht]
\centering
\includegraphics[width=12cm]{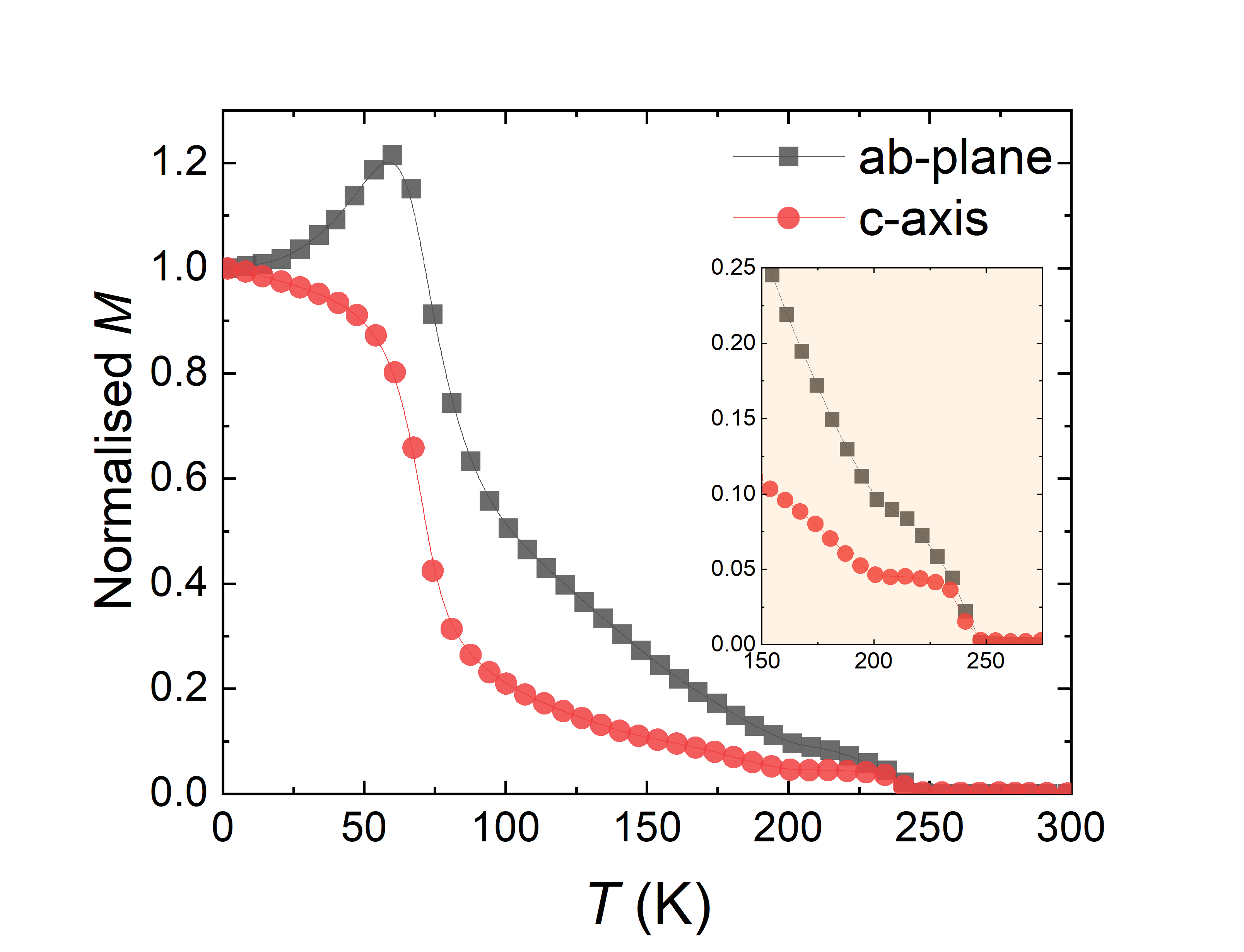}
\caption{$M$ vs $T$ curves in Na-CGT for two different orientations measured during cooling in the absence of a magnetic field.}
\label{SQUID_Supp_Fig1}
\end{figure}

As highlighted in the main text in Fig. 2(a), the $M$-$T$ curve in Na-CGT shows a different behaviour compared to CGT. In Figure~\ref{SQUID_Supp_Fig1}, the magnetisation response as a function of temperature is compared along the two orientations of out of plane (c-axis) and in-plane (ab-plane) for the Na-CGT sample. It is evident from this measurement that the magnetic ordering temperature has been enhanced and the value of $M$ goes to zero in the vicinity of 240\,K (inset of Fig.~\ref{SQUID_Supp_Fig1}).

\begin{figure}[ht]
\centering
\includegraphics[width=18cm]{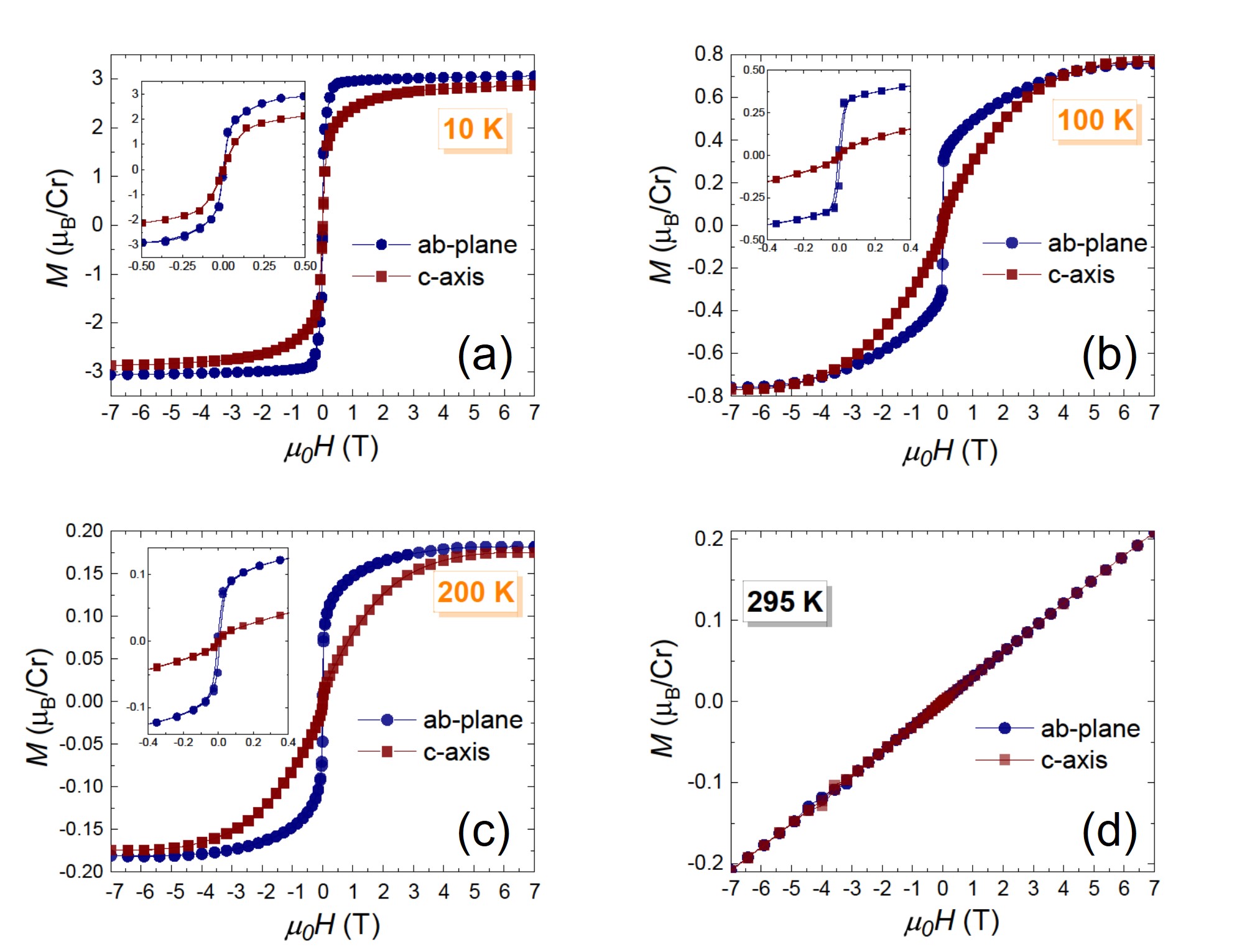}
\caption{$M$-$H$ loops in Na-CGT at different temperatures of (a) 10\,K, (b) 100\,K, 200\,K and (d) 295\,K. Inset in (a), (b) $\&$ (c) zooms in close to 0-magnetic field position.}
\label{SQUID_Supp_Fig2}
\end{figure}

In Fig.~\ref{SQUID_Supp_Fig2}, the dependence of magnetisation as a function of external magnetic field ($M$-$H$ curves) is shown for various different temperatures. Figure~\ref{SQUID_Supp_Fig2} (a) shows $M$-$H$ curves for Na-CGT at 10\,K for the ab-plane and c-axis orientations. Along the ab-plane, a saturation magnetic moment of 3.05\,$\mu_{\textrm{B}}$/Cr is obtained whereas along the c-axis it is found to be 2.87\,$\mu_{\textrm{B}}$/Cr. It is clear from the experimental $M$-$H$ curves that it requires less external magnetic field for the magnetisation to saturate along the ab-plane orientation than c-axis. Hence, it is evident from these measurements that a change in the magnetic anisotropy takes place after intercalation of Na-ions into CGT, and the spins prefer to align in the in-plane orientation compared to the out of plane orientation.  

Figure~\ref{SQUID_Supp_Fig2} (b $\&$ c) show the $M$-$H$ curves of Na-CGT at the elevated temperature of 200\,K where a clear magnetic response is observable, together with the uniaxial magnetic anisotropy. The magnetic response disappears at 295 K as shown in Fig.~\ref{SQUID_Supp_Fig2}(d)  where a linear trend is observed. This confirms that the ferromagnetic ordering phase transition to a paramagnetic phase takes place in Na-CGT.

\begin{figure}[ht]
\centering
\includegraphics[width=16cm]{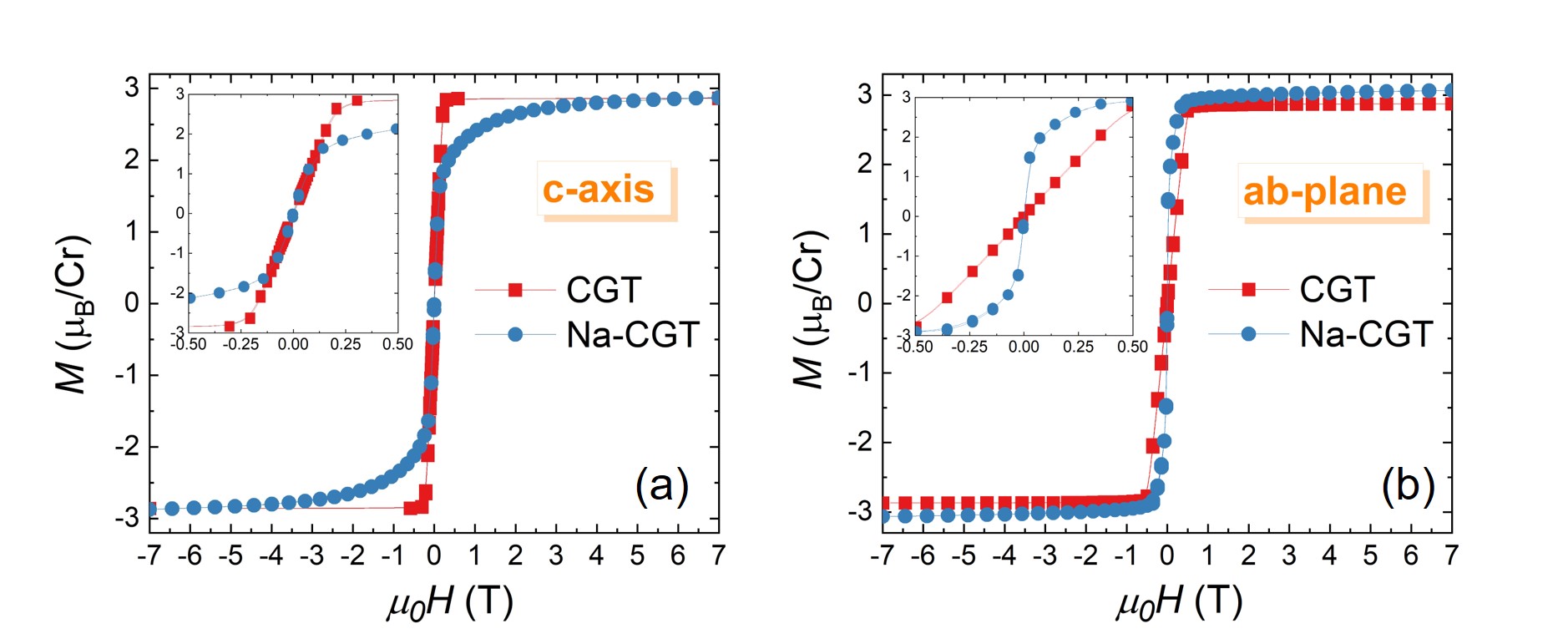}
\caption{$M$-$H$ loop Comparison between CGT and Na-CGT for two orientations, (a) c-axis and (b) ab-plane, at 10\,K. }
\label{SQUID_Supp_Fig3}
\end{figure}

Finally, in Figure~\ref{SQUID_Supp_Fig3}, $M$-$H$ curves for CGT and Na-CGT for two different orientations are compared at 10\,K. It is clearly visible that the magnetic state of CGT is modified post intercalation and a change in magnetic anisotropy takes place.

\clearpage

\section{Magnetic Susceptibility}

\begin{figure}[hb!]
\centering
\includegraphics[width=18cm]{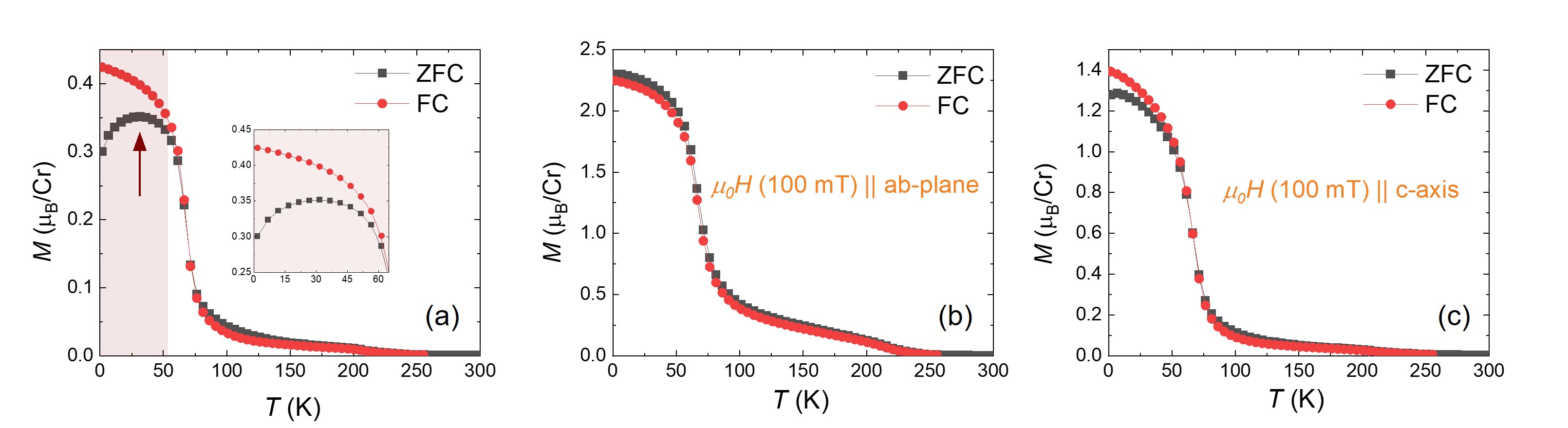}
\caption{(a) $M$ vs $T$ curves of Na-CGT for different measurement protocols with external magnetic field along the c-axis orientation, inset shows the zoomed-in divergence at low temperatures. (b) $\&$ (c) show that the bifurcation features start to disappear in both different orientations as the magnetic field is increased.}
\label{SpinGlass_S1}
\end{figure}

The temperature dependence of the magnetisation in Na-CGT with small external magnetic field (20\,mT) for the out of plane orientation is shown Figure~\ref{SpinGlass_S1} (a) and the bifurcation of magnetisation is also seen here with different field cooling protocols (similar to ab-plane orientation shown in the main text. Figure~\ref{SpinGlass_S1} (b) $\&$ (c) show $M$-$T$ curves for external field of 100\,mT and this highlights that as the magnetic field is increased from 20\,mT to higher field values, the branching between zero field-cooling (ZFC) and field-cooling (FC) protocols starts to disappear.

\begin{figure}[b!]
\centering
\includegraphics[width=10cm]{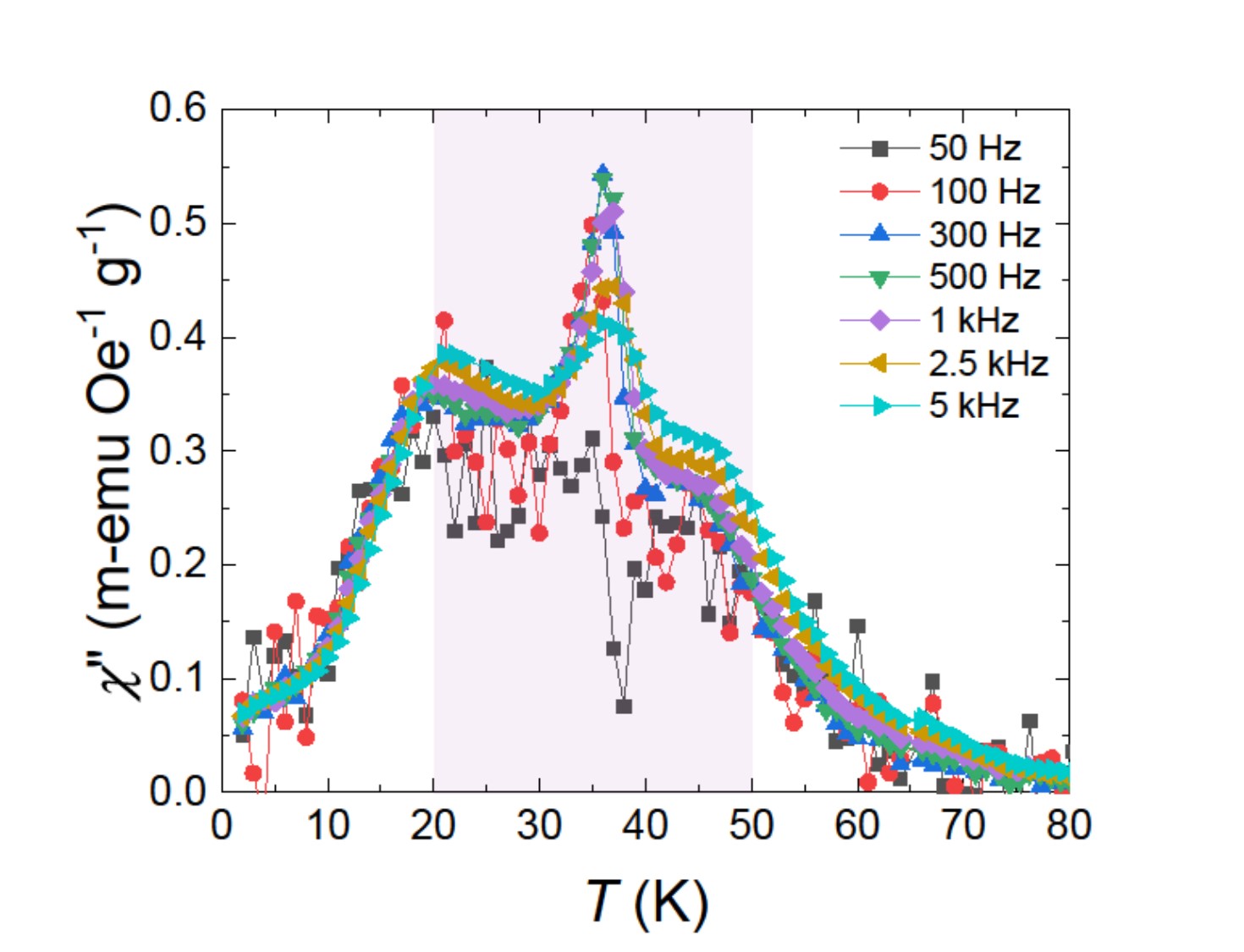}
\caption{Temperature dependence of the imaginary part of the ac susceptibility for different excitation frequencies in Na-CGT.}
\label{SpinGlass_S2}
\end{figure}

As mentioned in the main text, the ac magnetic susceptibility is expressed in complex form as $\chi_{a.c.} = \chi' + i \chi"$, where the real part ,$\chi'$, is the in-phase component and the imaginary part, $\chi"$, is the out of phase component~\cite{Mydosh1993SpinGlassBook_S}. Figure~\ref{SpinGlass_S2} shows the temperature dependence of $\chi"$ for different excitation frequencies, and even though the amplitude is smaller compared to the real part shown in Fig. 3(b) in the main text, the anomaly feature is clearly present in $\chi"$, supporting the spin-glass picture. In typical spin-glass systems, relative low amplitude of $\chi"$ is suggestive of distribution of relaxation mechanism instead of a single relaxation time~\cite{Balanda2013ACReview_S}. \\

\begin{figure}[ht]
\centering
\includegraphics[width=10cm]{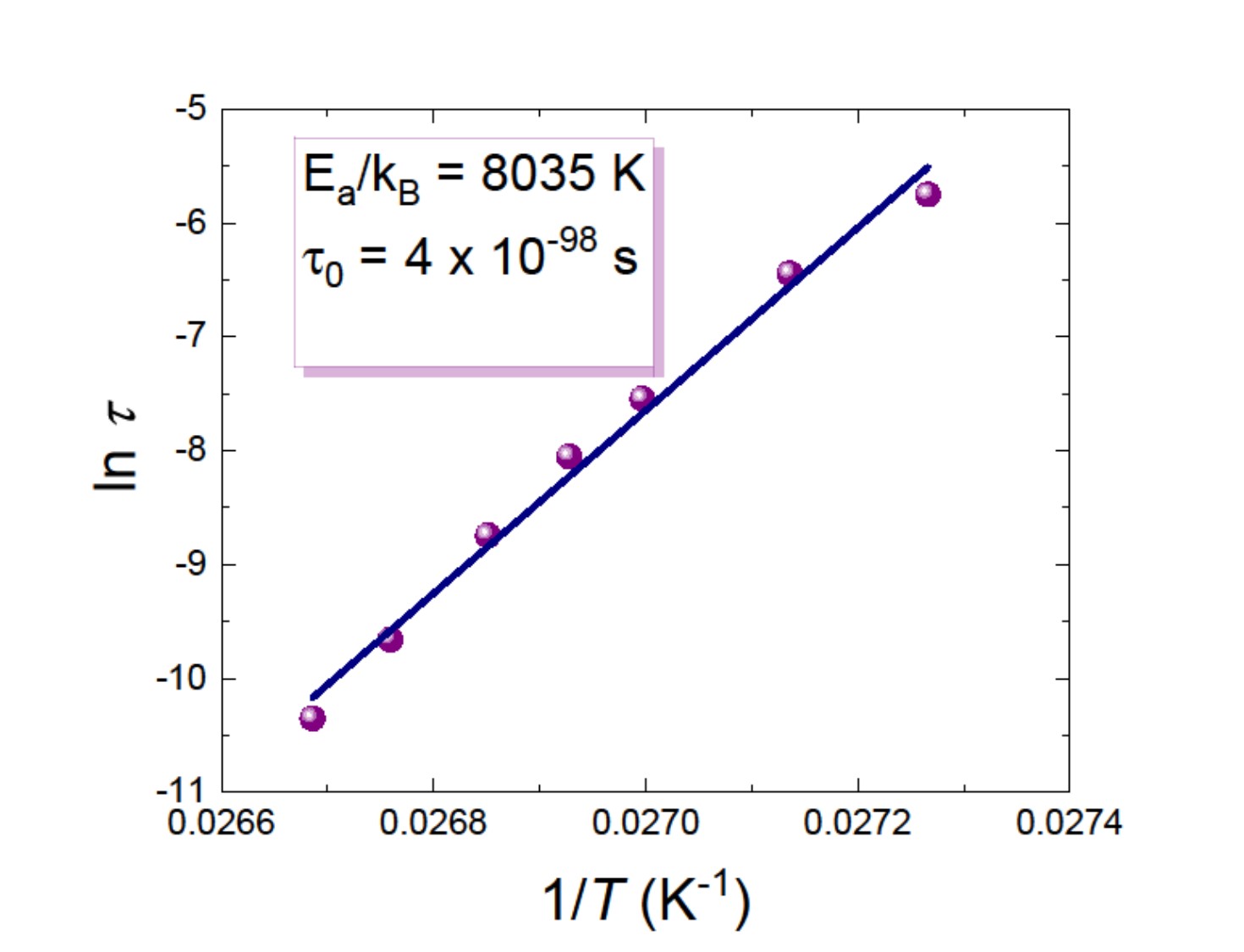}
\caption{Arrhenius Law fitting to the peak-maximum dependence on excitation frequency in the real part of ac susceptibility, using Eq.~\ref{eq:Arr}. The best fit parameters are shown in the figure. }
\label{SpinGlass_S3}
\end{figure}

Figure~\ref{SpinGlass_S3} shows the ln$\tau$ vs 1/$T$ plot of extracted from the peak-maximum $T_\text{g}$ in $\chi'$ measurements. We used the following Arrhenius-type equation~\cite{Mydosh1993SpinGlassBook_S,Balanda2013ACReview_S} to extract the characteristic activation energy $E_a$ as well as timescale $\tau$ given by:
\begin{equation}\label{eq:Arr}
\tau = \tau_{0} \hspace{0.2cm} \textrm{exp}\left[\frac{E_{a}}{k_{B}[T_{g}(f)]}\right]
\end{equation}
where $\tau = 1/2\pi f$ and $k_B$ is Boltzmann constant. As pointed out in the main text, even though the experimental data can be fit using Eq.(\ref{eq:Arr}), the best fit parameters of $E_\text{a}/k_\text{B}$ = 8035 K and $\tau_0 = 4 \times 10^{-98}$ are both unphysical. The relaxation mechanisms in Na-CGT can thus not be described in terms of non-interacting spins, suggesting the interacting nature of Na-CGT's magnetic clusters~\cite{Balanda2013ACReview_S}. \\

The Casimir-du Pr\'e equations analysis method is utilised as mentioned in the main text~\cite{Mydosh1993SpinGlassBook_S,ToppingBlundell2018ACReview_S}. These equations describe the dependence of both real and imaginary parts of the ac susceptibility on the excitation frequency at a given temperature, and are used in the description of slow magnetic relaxation that takes the form of a Debye model. In the Debye model, a system relaxes for an entity (polarisation or magnetisation) and assumes a non-interacting nature between the components of that entity. However, as shown above for Na-CGT, this entity belongs to clusters with finite interaction strength which makes it difficult to describe the experimental data by a Debye model and leads to the system having a distribution of relaxation times due to the presence of cooperative effects. A modification of the Casimir-du Pr\'e equations is carried out to take into account the broad distribution of relaxation times, referred to as the generalised Debye model~\cite{ToppingBlundell2018ACReview_S}. A detailed description of this model can be found elsewhere~\cite{ToppingBlundell2018ACReview_S}. This introduces a parameter, $0 < \alpha < 1$, which determines the spread of relaxation times, with two extremes of zero being the case for single relaxation time and the unity being for a broad distribution of relaxation times. The frequency dependence of ac susceptibility, ($\omega = 2\pi f$), in the generalised Debye model takes the following form~\cite{ToppingBlundell2018ACReview_S}:
\begin{equation}
 \chi(\omega) = \chi_S + \frac{(\chi_{0} - \chi_S)}{1+ (i\omega\tau)^{1-\alpha}}
\end{equation}
where $\chi_{0}$ represents the isothermal susceptibility and $\chi_{S}$ is the adiabatic susceptibility~\cite{ToppingBlundell2018ACReview_S}. The Casimir-du Pr\'e equations are modified by considering the following distribution function of $\tau$:
\begin{equation}\label{eq:disF}
g(\tau) = \frac{1}{2\pi}\frac{\sin(\alpha\pi)}{\cosh[(1-\alpha)\ln(\tau/\tau_c)]-\cos(\alpha\pi)}.
\end{equation}
\begin{figure}[t!]
\centering
\includegraphics[width=18cm]{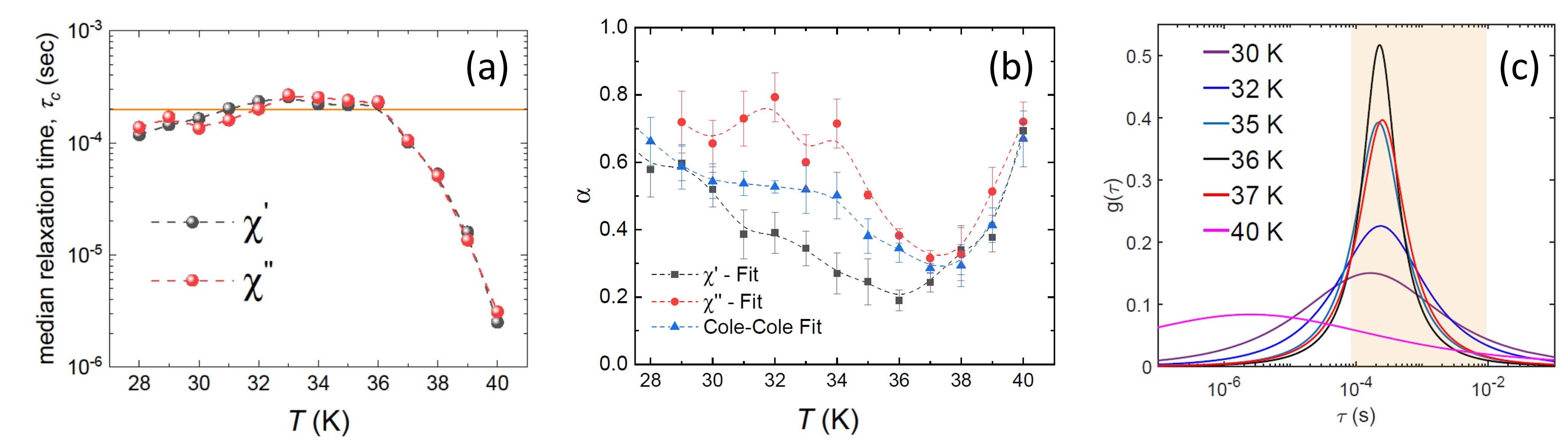}
\caption{(a) Temperature dependence of extracted median relation time $tau_c$ obtained from fit procedure carried out in Fig. 4 (a) $\&$ (b) in the main text. (b) $\alpha$ extracted by fitting experimental data by three different approaches, i.e. individual fits to each susceptibility component and Cole-Cole plot fitting. (c) Distribution function of $\tau$ calculated by \ref{eq:disF} with $\alpha$ for each temperature.}
\label{SpinGlass_S4}
\end{figure}
As such, the real and imaginary parts of the complex ac susceptibility become:
\begin{equation}
\chi' (\omega) = \chi_S + \frac{(\chi_{0} - \chi_S)}{2}\left[1 - \frac{\sinh[(1-\alpha)\ln(\omega\tau_c)]}{\cosh[(1-\alpha)\ln(\omega\tau_c)]+\cos[(1/2)(1-\alpha)\pi]}      \right] 
\end{equation}
\begin{equation}
 \chi'' (\omega) =  \frac{(\chi_{0} - \chi_S)}{2}\left[\frac{\sin[(1/2)(1-\alpha)\ln(\omega\tau_c)]}{\cosh[(1-\alpha)\ln(\omega\tau_c)]+\cos[(1/2)(1-\alpha)\pi]}      \right],
\end{equation}
where $\tau_c$ is the average relaxation time. We use these equations to fit the experimental data of the real and imaginary parts of ac susceptibility (e.g. those shown in Fig. 4). 
The temperature dependence of $\tau_c$ and $\alpha$ is extracted by the fitting and shown in Figs.~\ref{SpinGlass_S4} (a)-(b) respectively. There is a clear increase in the average relaxation time in the vicinity of 35\,K. This coincides with the earlier findings of bifurcation of magnetisation in different field-cooling protocols and the shift of peak-maximum in ac susceptibility measurements and it signifies slower dynamics as the freezing of the clusters takes place. The Cole-Cole analysis can also enables to determine $\alpha$ by the equation: 
\begin{equation}
\chi^{''}(\chi^{'}) = - \frac{(\chi_{0} - \chi_S)}{2\tan[(1-\alpha)(\pi/2)]} + \sqrt{(\chi^{'} - \chi_{S})(\chi_{0}-\chi^{'}) + \frac{(\chi_{0}-\chi^{'})^{2}}{4(\tan[(1-\alpha)(\pi/2)])^{2}}}.
\end{equation}
$\alpha$ extracted by the three different approaches, i.e. $\chi'$, $\chi"$ and Cole-Cole analysis, all show qualitatively similar temperature-dependent trends with comparable values, as shown in Fig.~\ref{SpinGlass_S4}(b). The $\alpha$ values between 0.2 to 0.8 further support the picture of multiple channels of the spin-glass relaxation presented in the main text. Finally, we show the calculated $g(\tau)$ by extracted $\alpha$ from the Cole-Cole analysis in Fig.~\ref{SpinGlass_S4}(c). This visualises the statistical distribution of relaxation times for magnetic clusters in Na-CGT.\\

\section{Magneto-transport}

\begin{figure}[ht!]
\centering
\includegraphics[width=18cm]{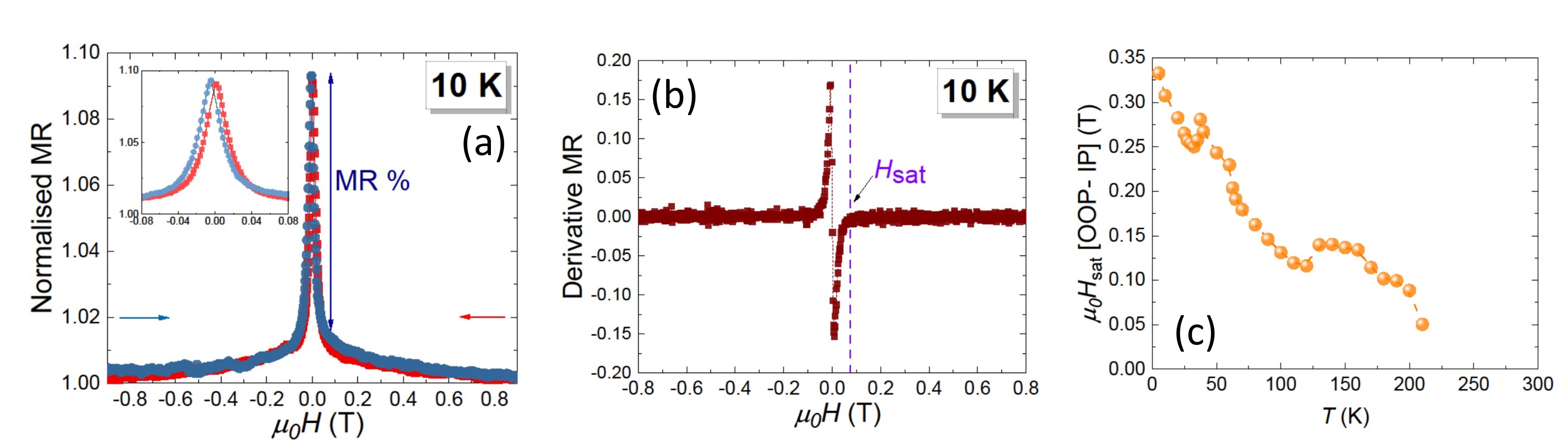}
\caption{(a) Typical MR measurements of Na-CGT measured during the magnetic field sweep for 10 K. Resistance is normalised at the high field limit for each scan. The inset highlights the MR behavior close to zero-field, showing a small hysteresis. (b) Plot of derivative values of MR for determining the saturation field $H_\text{sat}$ from each scan. (c) Temperature dependence of the difference in $H_\text{sat}$ for two orientations in Na-CGT to represent the temperature dependence of the uniaxial magnetic anisotropy.}
\label{MR_Supp_Fig1}
\end{figure}

Here we show technical details and further analysis of magnetoresistance(MR) measurements measured in our Na-CGT. Figure~\ref{MR_Supp_Fig1}(a) shows the MR measurements for Na-CGT at 10\,K while the magnetic field is swept from positive to negative and then negative to positive directions. A peak-shaped resistance change is observed upon sweeping the external magnetic field around zero. The MR value plotted in Fig. 2(f) is deduced by the relative resistance difference between the high field and zero field values. We believe that this is anisotropic magnetoresistance where the resistance change depends on the angle between the current direction and magnetic moment orientation. On this assumption, the resistance represents the moment orientation and can be used for analysis of magnetic switching behaviour. In order to more reliably identify the completion of magnetisation switching, we use the derivative form of MR as shown in Fig.~\ref{MR_Supp_Fig1}(b). This approach is also effective to remove any contribution from linear MR components.  

The magnitude of the uniaxial magnetic anisotropy can be estimated by the difference in saturation fields $H_\text{sat}$ for the out of plane (OOP) and in-plane (IP) orientations, as shown in Figure~\ref{MR_Supp_Fig1} (c). The finite $H_\text{sat}$ difference representing the presence of magnetic anisotropies is another proof of the presence of long-range magnetic order. The temperature dependence is good agreement with the $T_\text{C}$ estimated from different measurement techniques such as magnetometry. 
%The values of A small dip in anisotropy is also observed close to the signature of the cluster glass behaviour seen in ac susceptibility measurements. Overall, magneto-transport measurements on bulk Na-CGT samples show a drastic change in the magnetic behaviour compared to the undoped CGT case. It also confirms increase in the Curie temperature, a change in magnetic anisotropy in the new system, and even though it is a dc measurement probe, it somewhat captures a small change in anisotropy close to the cluster glass observed in the low temperature phase. 

\section{\rrev{Theoretical Methods}}\label{sec4}
\subsection{\rrev{Ab initio simulations}}
\rrev{To provide additional insights of the spin-glass states in Na-CGT systems, we undertook multi-scale simulations using density functional theory (DFT) to calculate the magnetic parameters (e.g., exchange interactions, anisotropies, etc.), and Monte Carlo to simulate variations of Curie temperature. On the DFT part, geometry optimizations of bulk Cr$_{2}$Ge$_{2}$Te$_{6}$ are performed using Vienna Ab initio Simulation Package (VASP)\cite{Kresse1999,Kresse1994}. The Na intercalation is performed by changing the position of Na in the unit cell and the magnetic parameters are evaluated for the lowest energy configuration. The generalized gradient approximation (GGA) has been used to treat the exchange-correlation interactions within the Perdew-Burke-Ernzerhof (PBE) form\cite{Perdew1996}. We use 18$\times$18$\times$6 Monkhorst-Pack $k$-point mesh in our calculations for Brillouin zone (BZ) integration.\cite{Monkhorst1976} The lattice parameters and atomic coordinates are optimized by the energy minimization technique based on the conjugate gradient algorithm with a force component tolerance of 0.01 eV/\AA\ on each atom. The cutoff for the energy of plane-wave basis set is considered to be 500 eV.}

\rrev{The full-potential linear muffin-tin orbital (FP-LMTO) method, implemented in the RSPt code\cite{Wills2000} is used to study the magnetic and electronic properties of Fe$_{5-\delta}$Ni$_{\delta}$GeTe$_{2}$ monolayers. First we perform standard density functional theory (DFT) calculations, The converged DFT calculations are the starting point to perform our DFT+U calculations as implemented in RSPt\cite{Wills2000}. The value of Hubbard $U$ parameter or $U_\mathrm{eff}=U-J_\mathrm{H}$ ($J_\mathrm{H}$ is the Hund's exchange) for Cr-3$d$ states is used to be 0.5 eV, similar with  the previous studies\cite{Gong_Nature2017_S}.}

\subsection{\rrev{Calculation of exchange interactions}}
\rrev{The isotropic symmetric exchange interactions $J_{ij}$ are calculated within the full-potential linearized muffin-tin orbital (FPLMTO) basis implemented in the RSPt code. From the LMTO basis, one can construct the Bloch sums to solve the DFT eigenvalue problem and subsequently for the one-electron Green’s function.  }

\rrev{The $J_{ij}$ couplings are computed using force-theorem based Green's function formalism\cite{Yaroslav_2015}. The intersite exchange parameters are calculated using eq.~\ref{Jij_DFT}:}

\begin{figure*} [t]\centering  
\includegraphics[width=0.75\textwidth]{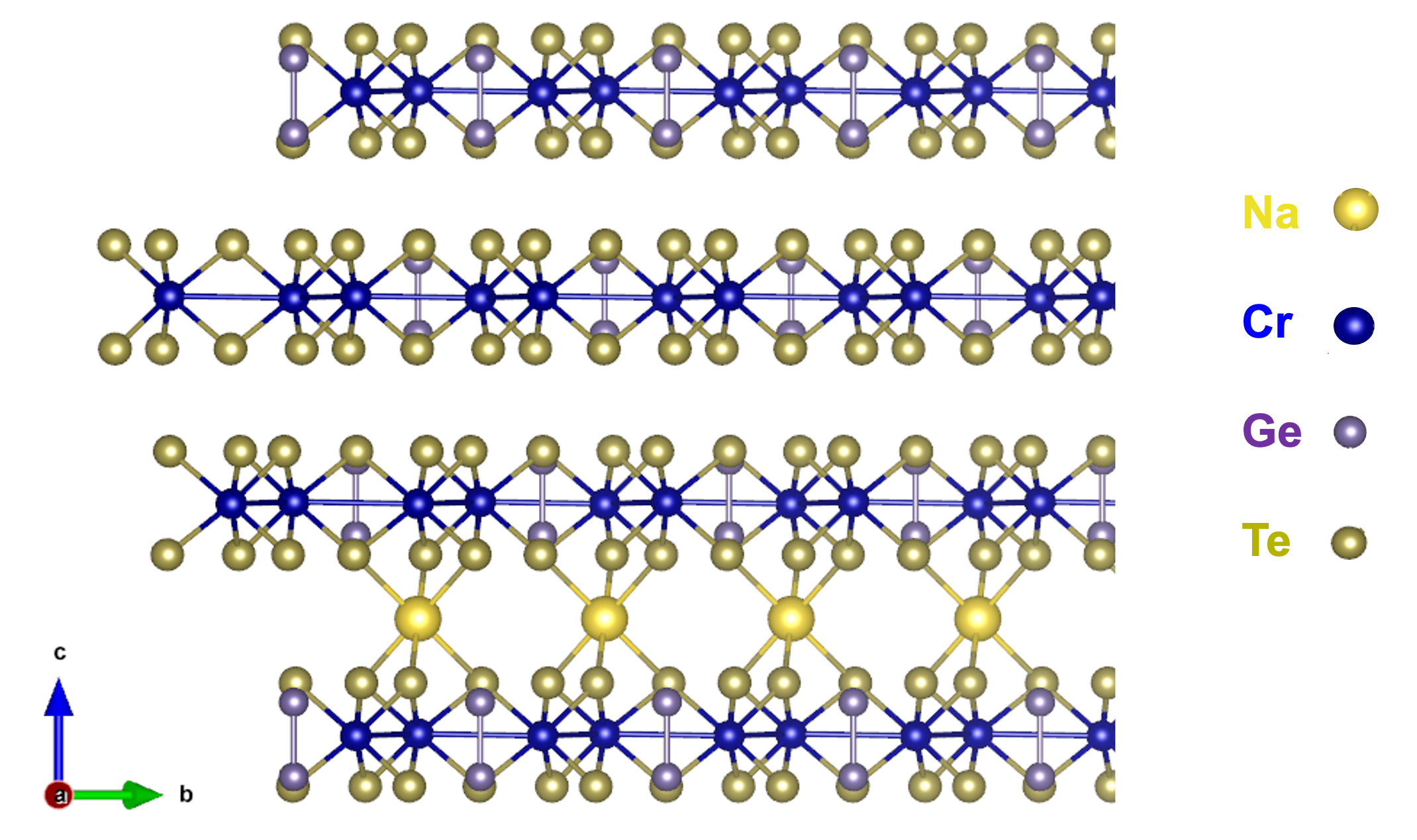}
\caption{Schematic of the Na atoms intercalating Cr$_{2}$Ge$_{2}$Te$_{6}$ 
at 16.67\% concentration. Atoms are colored following the labeling at the right side. In-plane ($ab$) and out-of-plane ($c$) directions are highlighted at the cartesian axis. 
%with binding energy -3.10 eV/cell.
}
\label{structure}
\end{figure*}
\begin{equation}
    \rrev{J_{ij} = \frac{T}{4}\sum_{n} [\hat{\Delta}_{i} (i\omega_{n})\hat{G}_{ij}^{\uparrow}(i\omega_{n})\hat{\Delta}_{j} (i\omega_{n})\hat{G}_{ji}^{\downarrow} (i\omega_{n})],
    \renewcommand{\theequation}{S\arabic{equation}}}
    \label{Jij_DFT}
\end{equation}
\rrev{where $\Delta_{i}$ and $\hat{G}_{ij}$ are the onsite spin splitting and the spin-dependent intersite Green's function, respectively. The trace in eq.~\ref{Jij_DFT} is over the orbital degrees of freedom. $\omega_{n}=2\pi T(2n+1)$ and $T$ are the $n$th fermionic Matsubara frequency and the temperature, respectively. The onsite exchange splitting term $\Delta_{i}$ includes the self-energy, which is given by:}
        \begin{equation}
            \rrev{\Delta_{i}(i\omega_{n})=H_{KS}^{\uparrow}+\Sigma_{i}^{\uparrow}(i\omega_{n})-H_{KS}^{\downarrow}-\Sigma_{i}^{\downarrow}(i\omega_{n}),
            \renewcommand{\theequation}{S\arabic{equation}}}
            \label{Delta}
        \end{equation}
\rrev{where, $H_\mathrm{KS}$ and $\Sigma_{i}$ are the Kohn-Sham Hamiltonian and site-dependent self-energy. The self-energy is obtained by solving the DMFT equations. In DMFT calculations, The frequency-dependent self-energy is obtained from DMFT. The exchange splitting also depends on frequency\cite{Yaroslav_2015}.} 

 \rrev{In the fully-relativistic case the generalized Heisenberg model becomes:}
\begin{equation}
    \rrev{H = -\sum_{i\neq j} e_{i}^{\alpha}J_{ij}e_{j}^{\beta},
    \renewcommand{\theequation}{S\arabic{equation}}}
    \label{Jij_relativistic}
\end{equation}
\rrev{where $\alpha$, $\beta = x, y, z$. The magnetic exchange parameters $J_{ij}$ are ($3\times3$) tensors in the considered fully-relativistic case, which is a generalization of the $J_{ij}$ scalar parameters\cite{Yaroslav_2015}. To calculate the magnetic exchange interactions we use $21\times21\times1$ k-point mesh. }

\subsection{\rrev{Exchange interactions in Na-intercalated CrGeTe$_3$ magnets}}

\rrev{In the following we calculated the isotropic exchange interactions (Table \ref{Symm_J}), antisymmetric exchange interactions (e.g., Dzyaloshinskii-Moriya - $D$) (Table \ref{DMI}), and anisotropic exchange (Table \ref{lambda}). The inclusion of Na dopants break the symmetry of the Cr atoms, which generated additional interactions responsible for likely competition between pristine areas where ferromagnetism is present, and areas where Na-induced potential frustrations. Figure \ref{structure} shows a schematic of the CGT layers with intercalated Na atoms into the system. }
%Slightly variations are observed for different configurations.   

\begingroup
\begin{table*}[h]\centering
{
\caption{\rrev{Isotropic symmetric exchange interactions between in-plane ($J_{ab}$) and out-of-plane ($J_{c}$) atoms at different nearest-neighbors (NNs):  
$J_{ab1}$, $J_{ab2}$, $J_{ab3}$, $J_{ab4}$ for first, second, third and forth NNs, respectively; and, similarly, for out-of-plane interactions, $J_{c1}$, $J_{c2}$, $J_{c3}$, $J_{c4}$. Pristine and Na-intercalated CrGeTe$_3$ (CGT) bulk systems are considered. With the inclusion of Na atoms into CGT, there is a symmetry breaking at the Cr sites closest to the dopants which reflects in the variation of the exchange parameters relative to the areas far apart from the dopants. The first and second rows in the table for Na-intercalated correspond to the interactions between Cr atoms in the immediate vicinity of Na and away from Na, respectively. The $J$ values are in the units of meV, and the Na concentration is 16.67\%. }
}
%\centering
\begin{tabular}{|c|c|c|c|c|c|c|c|c|}
\hline
System  & $J_{ab1}$ & $J_{ab2}$ & $J_{ab3}$ & $J_{ab4}$ & $J_{c1}$ & $J_{c2}$ & $J_{c3}$ & $J_{c4}$\\ 
\hline
 Pristine & 5.478  & -0.019 & 0.687 & -0.052 & 0.121 & 0.302 & -0.068  & -0.053 \\
 \hline
 Na-intercalated & 5.325 & 0.790 & 1.667 & 0.227  &  -0.162 & 0.471 & 0.259  & 0.4689 \\
  & 4.369 & -0.171  &  1.024 & -0.297  &   &  &   & \\
\hline
\end{tabular}
\label{Symm_J}}
\end{table*}
\endgroup

\begingroup
\begin{table*}[h]\centering
{
\caption{\rrev{Similarly as in Table \ref{Symm_J}, but for 
antisymmetric exchange interactions (e.g., Dzyaloshinskii-Moriya) $D$. In pristine CGT, the magnitude of $D$ is only finite for second NNs ($D_{ab2}=0.159$ meV), which follows the symmetry of the honeycomb lattice. However, as Na atoms are included between the layers, several Cr atoms broken their immediate symmetry generating a finite magnitude of $D$ at different number of NNs (first row at Na-intercalated). For the Cr atoms away from Na, they behave similarly as in the pristine case but with slightly larger magnitudes of $D$ (second row at Na-intercalated).  }}
%\centering
\begin{tabular}{|c|c|c|c|c|c|c|c|c|}
\hline
System  & $D_{ab1}$ & $D_{ab2}$ & $D_{ab3}$ & $D_{ab4}$ & $D_{c1}$ & $D_{c2}$ & $D_{c3}$ & $D_{c4}$\\ 
\hline

 Pristine & 0.000 & 0.159 & 0.000 & 0.000 & 0.000 & 0.005 & 0.003 & 0.000  \\
 \hline
 Na-intercalated & 0.309 & 0.165 & 0.149 & 0.028  & 0.177  & 0.000 & 0.128  & 0.102 \\
  & 0.000 & 0.186  &  0.000 & 0.020  &   &  &   & \\
\hline

\end{tabular}
\label{DMI}
}
\end{table*}
\endgroup

\begingroup
\begin{table*}[h!]\centering
{
\caption{\rrev{Similarly as in Table \ref{Symm_J}, but for 
anisotropic symmetric exchange interactions $\lambda$. }}
%\centering
\begin{tabular}{|c|c|c|c|c|c|c|c|c|c|}
\hline
System  & $\lambda_{ab1}$ & $\lambda_{ab2}$ & $\lambda_{ab3}$ & $\lambda_{ab4}$ & $\lambda_{c1}$ & $\lambda_{c2}$ & $\lambda_{c3}$ & $\lambda_{c4}$\\ 
\hline

 Pristine &  0.192  & 0.003 & 0.036 & 0.008 & 0.000 & 0.008 & 0.005 & 0.002 \\
 \hline
 Na-intercalated  & 0.129 & 0.017 & 0.049 & 0.003  &  0.000 & 0.001 & 0.007 & 0.006 \\
  & 0.156  & 0.0178  & 0.0795  & 0.005  &   &  &   &   \\
\hline
\end{tabular}
\label{lambda}
}
\end{table*}
\endgroup

\subsection{\rrev{Magnetic Anisotropy}}
\rrev{The single-ion anisotropy is estimated based on the one-shot fully-relativistic calculations for the spin axis pointing along the $x$, $y$, and $z$ directions. These calculations are run starting from the fully-converged self-consistent non-relativistic electronic structure. The value of anisotropy is obtained from the difference in the sum of the energy eigenvalues for the three spin directions, as already mentioned. Note that, this approach, which is based on the force theorem, often is a more accurate way of determining the magnetic anisotropy, compared to relativistic total energies\cite{Vladi_PRM_2022}. We use the k-point grids with dimensions $48\times48\times16$ for the magnetic anisotropy calculations. A change in orientation from easy-axis to easy-plane is observed after the introduction of Na atoms. }

\begingroup
\begin{center}
\begin{table*}[h!]\centering
{
\caption{\rrev{Anisotropy energy $K$ for pristine CGT and Na-intercalated CGT. $K=E_{\perp}-E_{\parallel}$ in units of (meV/Cr), where $E_{\perp}$ and $E_{\parallel}$ correspond to perpendicular and parallel spin configurations to the surface, respectively. }}
\begin{tabular}{|c|c|c|c|c|c|c|c|c|c|}
\hline
System  & K\\ 
\hline
Pristine &  -0.08   \\
 \hline
 Na-intercalated  & 0.23 \\
\hline
\end{tabular}
\label{Symm_oper}
}
\end{table*}
\end{center}
\endgroup

\subsection{\rrev{Monte Carlo simulations}}

\rrev{Solving the Heisenberg spin Hamiltonian, considering localized spin moments, the Curie temperature $T_\mathrm{C}$ is computed performing Monte Carlo simulations: }

 \begin{equation}
  \rrev{H = -\sum_{i\neq j} J_{ij}\vec e_{i}{\cdot} \vec e_{j}-\sum_{i\neq j} \vec{D}_{ij}\cdot (\vec e_{i}{\times} \vec e_{j})-\sum_{i\neq j} K_{i}(e_{i}^{z})^{2}-\sum_{i\neq j}\lambda_{ij}(e_{i}^{z}e_{j}^{z}),}
  \renewcommand{\theequation}{S\arabic{equation}}
   \label{H}
\end{equation}
\rrev{where $J_{ij}$, $D_{ij}$ and $\lambda_{ij}$ are isotropic symmetric, antisymmetric and anisotropic symmetric interactions between $i$th and $j$th species, $K_{i}$ is the single-ion anisotropy. Such Hamiltonian has been intensively investigated\cite{Elton_QuantumRescaling2021,Srini23,Coronado24,Jenkins22,Yonathan23,Balicas23,Balicas24,Koperski24,Hicke22,Li24} throughout many systems and conditions which provide a robust background for the simulations.   }

\begin{figure*} [h]\centering  
\includegraphics[width=0.5\textwidth]{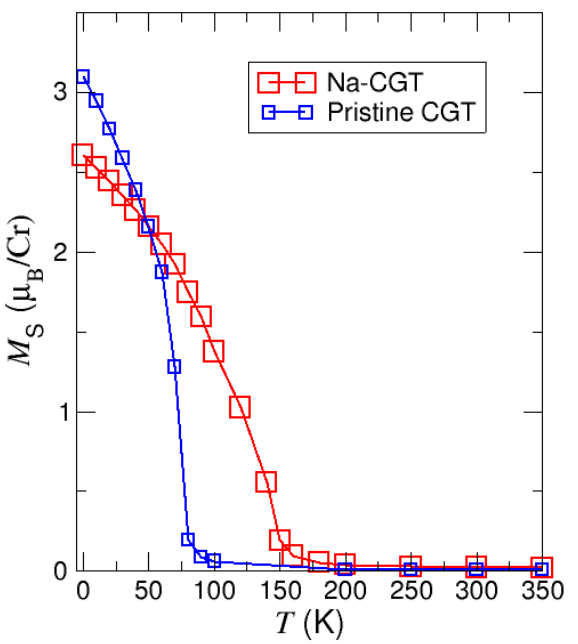}
\caption{\rrev{Magnetisation versus temperature for pristine CGT and Na-CGT with 16.67\% Na concentration. Estimation of the Curie temperature ($T_\text{C}$) resulted in $T_\text{C}\approx 69$ K and $\approx$150 K for CGT and Na-CGT, respectively.  }
%with binding energy -3.10 eV/cell.
}
\label{Tc}
\end{figure*}

\rrev{To estimate the Curie temperatures ($T_\text{C}$) shown in Figure \ref{Tc}, we performed classical Monte Carlo (MC) simulations via UppASD code\cite{UppASD}, where the calculated magnetic parameters are implemented in Eq.~\ref{H}. To achieve properly averaged properties, calculations are done for three ensembles in the supercell with size 30$\times$30$\times$10, where periodic boundary conditions are imposed along $x$ and $y$ axes. $T_\mathrm{C}$ is estimated by monitoring the cross sections of fourth-order cumulants of magnetization. The number of Monte Carlo steps considered in our calculations is $1\times10^{6}$.}
   
\makeatletter
\renewcommand\@biblabel[1]{S#1}
\makeatother

%\bibliography{sample}

\end{document}